\documentclass[lettersize,journal]{IEEEtran}  

\usepackage{graphicx, caption, subcaption}
\usepackage{amsmath,amsthm,amssymb}
\usepackage{hyperref}

\newcommand{\Position}{\mathbf{p}}
\newcommand{\DeltaRef}{\delta_{\mathrm{ref}}}
\newcommand{\Chl}{chlorophyll~\textit{a}} 

\title{\LARGE \bf Adaptive Sampling of Algal Blooms \\Using Autonomous Underwater Vehicle and Satellite Imagery:\\ Experimental Validation in the Baltic Sea}
\author{Joana Fonseca*, Sriharsha Bhat, Matthew Lock, Ivan Stenius, Karl H. Johansson   
\thanks{*Corresponding author.}
\thanks{This work is supported by Knut and Alice Wallenberg Foundation, Swedish Research Council, and Swedish Foundation for Strategic Research.} 
\thanks{J. Fonseca, M. Lock, and K. H. Johansson are with the Division of Decision and Control Systems, School of Electrical Engineering and Computer Science, and Digital Futures. S. Bhat and I. Stenius are with the Department of Aeronautical and Vehicle Engineering, School of Engineering Sciences. KTH Royal Institute of Technology, SE-100 44 Stockholm, Sweden.
\texttt{\{jfgf, svbhat, mwlock, stenius, kallej\}@kth.se}.}}

\begin{document}
\maketitle

\begin{abstract} 
This paper investigates using satellite data to improve adaptive sampling missions, particularly for front tracking scenarios such as with algal blooms.
Our proposed solution to find and track algal bloom fronts uses an Autonomous Underwater Vehicle (AUV) equipped with a sensor that measures the concentration of {\Chl} and satellite data.
The proposed method learns the kernel parameters for a Gaussian process (GP) model using satellite images of {\Chl} from the previous days. Then, using the data collected by the AUV, it models {\Chl} concentration online. We take the gradient of this model to obtain the direction of the algal bloom front and feed it to our control algorithm.
The performance of this method is evaluated through realistic simulations for an algal bloom front in the Baltic~sea, using the models of the AUV and the {\Chl} sensor. 
We compare the performance of different estimation methods, from GP to curve interpolation using least squares.
Sensitivity analysis is performed to evaluate the impact of sensor noise on the methods' performance.
We implement our method on an AUV and run experiments in the Stockholm archipelago in the summer of 2022.
\end{abstract}


\section{Introduction}

Researchers worldwide have been using satellites, remote sensing, and buoys to gather information on ocean phenomena and inform forecast models. 
These methods tend to be expensive, inefficient for spatial coverage, or unable to provide trustworthy data. They always have a human in the loop for decision-making or data post-processing.
With this paper, we design and implement a solution and make the open-source software available, which allows for autonomous coastal surveys using autonomous underwater vehicles (AUVs), focusing on front tracking of ocean phenomena such as algal bloom fronts.

\subsection{Background and Motivation}

Fronts are boundaries between water masses that differ significantly in the value of one or more variables, such as temperature, salinity, or substance concentrations. 
These fronts shape marine ecosystems as their presence indicates the occurrence of several physical and biological processes of interest, including transition zones, jets, eddies, and phytoplankton blooms~\cite{cornillion07,southern_fronts}. 
Among these frontal phenomena, harmful algal blooms (HABs) are the motivating scenario of this paper. 
According to \cite{Harmful}, ``HABs cause human illness, large-scale mortality of fish, shellfish, mammals, and birds, and deteriorating water quality''.
HABs occur when algae colonies experience abnormal growth, which results in the production of harmful toxins~\cite{NOAA}.
These toxins can cause significant harm to marine ecosystems and pose a danger to human activities in the Baltic Sea, such as tourism and aquaculture.
Accurate information about the location and movement patterns of algal blooms is crucial to monitor and mitigate these detrimental effects.  
Traditional methods for observation, such as satellite imaging or ship-towed sensors, are generally unable to provide measurements at the spatial and temporal resolutions required to understand dynamic ocean phenomena \cite{munk}. 
While remote sensing with satellites can offer a broad overview, such data is weather-dependent and prone to false positives in coastal areas. 
Thus, there is a significant scientific and societal interest in obtaining \textit{in~situ} measurements and developing systems for automated monitoring.

\begin{figure}[btp]
\centering
\includegraphics[width=.49\textwidth,trim={115mm 60mm 95mm 60mm},clip]{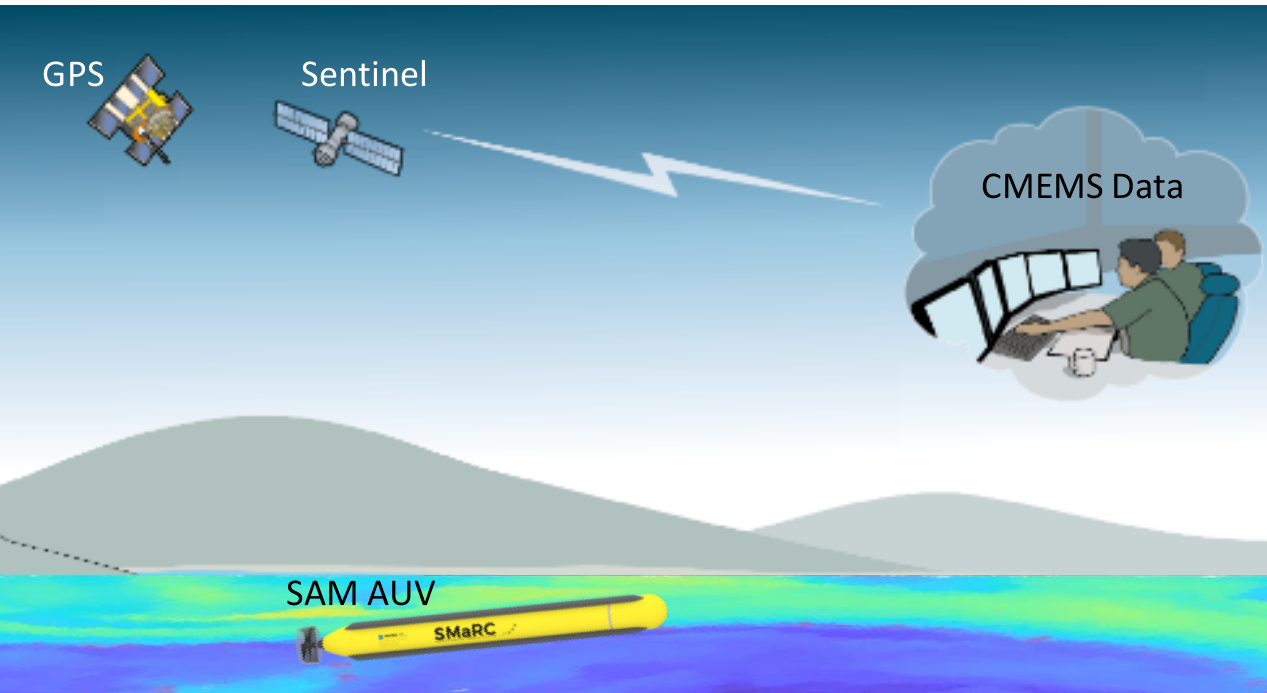}
\caption{System overview with the SAM AUV, the GPS, the Sentinel satellite, and the CMEMS data.} \label{overview}
\vspace{-5mm}
\end{figure}

This paper proposes an approach to adaptive sampling of algal bloom fronts using an AUV informed by satellite imagery.
The proposed algorithm enables tracking multiple water parameters, including {\Chl} concentrations, turbidity, dissolved oxygen, and salinity. 
Out of these broad sampling possibilities, we are particularly interested in chlorophyll fronts due to their connection with the occurrence of harmful algal blooms (HABs) in the Baltic~\cite{Wasmund2002}.
In Fig.~\ref{overview}, we illustrate this cyber-physical system, consisting of the AUV, the GPS signal that it uses for localization, the Sentinel satellite, which provides raw imagery of the region, and the Copernicus Marine Environment Monitoring Service (CMEMS), which reanalyses the imagery from the Sentinel satellite into more accurate datasets that can then be used to inform our AUV.

\subsection{Related work}

There has been significant work in developing solutions for autonomous ocean sampling.
These solutions involve satellites, research vessels, floats, gliders, or AUVs.
Satellite imagery offers broad datasets with varying resolutions but falls short for shallow waters.
Despite the shortcomings of satellite imagery for coastal data, alternate solutions for near-shore regions have been the topic of extensive research for the past 50 years in efforts to bridge that gap \cite{review_satellite}. 
For example, in \cite{satellite_liu}, the authors propose a novel wavelet analysis with satellite data from repeating paths for coastal coverage. 
On the other hand, several solutions for \textit{in~situ} sampling, such as research vessels, generally use towed platforms integrating cameras, sensors, acoustic devices, and sidescan data \cite{vessel_purser}.
In some cases, they also use underwater electronic holographic cameras for studying marine organisms such as plankton \cite{holographic_camera}. 
Focusing now on autonomous \textit{in~situ} sampling, one of the most widely used platforms is the profiling float, which moves with underwater currents and is sometimes referred to in the literature as a Lagrangian float.
Profiling floats are more controllable in regions of more diverse currents, but controlling the horizontal motion of a profiling float remains challenging \cite{controlability_floats}.
Despite their limited controllability, profiling floats such as in the Argo Program are perhaps the greatest international collaborative effort in the history of oceanography and provide researchers with open access to comprehensive data sets \cite{argo_program}.
Underwater gliders are more controllable than floats; in \cite{glider_Jeff}, the authors who built the glider "Spray" define gliders as \textit{autonomous profiling floats that use a buoyancy engine to move vertically and wings to glide horizontally}.
The creators of the glider "Slocum" stated that their glider originated from the idea of adding horizontal propulsion to floats \cite{glider_Douglas}.
The "Seaglider" \cite{glider_charles} is another example of one of the first gliders to be developed.
With the development of gliders, the problem of under-actuated controllability has been a subject of analysis \cite{glider_naomi}.
Gliders fall short in deterministic controllability, usually have a very limited sensor payload, and require a minimum operating depth of approximately 50m. These make them unsuitable for coastal applications in shallow water. 
AUVs, on the other hand, can carry different payload sensors, have better controllability, and can operate at multiple depths.

AUVs have been considered for solving the environmental sampling problem. 
One of the most common solutions in the literature is the open-loop scenario with a fixed sampling pattern. 
The most widely used sampling pattern is, undoubtedly, the lawn-mower \cite{chavez1997moorings}, which has been used for both single-agent \cite{willcox2001performance} and multi-agent scenarios \cite{das2012coordinated}.
However, other relevant methods, such as the spiral and circular patterns in \cite{ozer}, aim to improve efficiency and robustness.
Or the oval spiral coverage strategy to plan coverage paths that better suit oval-shaped areas of interest \cite{path_coverage}.
While these open-loop strategies enable and even guarantee coverage of survey areas, they are inherently not designed to react or respond to changes in the observed features. In such cases, there is great motivation to close the loop. 

Over the past two decades, there has been significant effort in closing the loop using adaptive sampling strategies.
Adaptive sampling is a closed-loop control architecture in which an agent autonomously makes decisions during a mission in response to environmental changes.
As reviewed in \cite{hwang2019auv}, adaptive sampling can be divided into three distinct objectives: source localization, front determination, tracking, and mapping. These objectives can be realized for different types of targets: thermoclines, algal blooms, oils spills, etc., using different vehicle configurations: single-agent, multi-agent with leader-follower, cooperative multi-agent, etc.
In \cite{thermocline}, the focus is on covering multiple thermoclines as they evolve in time and space in a dynamic water column.
In \cite{peter2004} and \cite{fiorelli2006multi}, the problem of source localization using a multi-agent system is approached as a gradient climb with optimal formation to minimize the gradient estimate error.
An example of front determination is \cite{Fossum_adaptive}, where the authors use a single AUV to find and track a salinity and temperature front while zig-zagging around it.
Considering the mapping problem, in \cite{namik_mapping}, the authors evaluate how to find the optimal path to maximize the accuracy of the field estimates for single and multi-agent scenarios.
Also, in \cite{john_leonard}, the front determination problem is considered with a single AUV zig-zagging the front, but for bathymetric contours.
In \cite{yanwu_layer}, a deep Chlorophyll maximum layer is tracked and mapped using three agents moving adaptively - one on the surface, one tracking the layer, and one mapping the area around the layer.
There are other examples similar to our objective in this paper, in which an adaptive sampling algorithm is proposed that augments a standard Gaussian process (GP) with a nearest neighbor prior \cite{McCammon}. Unlike our approach, this paper does not use external data to aid the vehicle's decisions, while being similar to our method of building a GP model from measurements.

In this field, one of the rising problems considers integrating external data (satellite imagery, numerical models and predictions, etc.) with local measurements taken by autonomous agents such as AUVs.
Specifically, the problem is how to aid ocean sampling missions using external data.
An early example uses a predictive ocean model to assist in motion planning for steering an AUV to a high-valued location \cite{ryan2010}. 
Here, it is assumed that there is a predicted model for the day of the mission, which is not available in this paper's scenario.
Other related results include using knowledge from previous missions to create a model \cite{Das2015} and using hydrodynamic and biological model systems as prior information \cite{Fossum_GP}.
Such adaptive sampling is closely linked to data assimilation. 
Among the first works in data assimilation for oceanic applications are \cite{dickey2003emerging,schofield2002long}. 
In \cite{Fossum_phytoplankton}, the data collected by the AUV is augmented with remote sensing, buoys, a ship, particle imaging systems, and discrete water samples.
In \cite{heaney2007nonlinear}, the authors use a genetic algorithm to optimize the deployment, measurements, and information gain of a team of AUVs, ASVs, and mooring platforms. 
Such measurements are then included in an estimation framework to estimate and forecast environmental parameters given a dynamic ocean model.
In \cite{chang2019data}, the authors propose using generic environmental models that are updated with data collected by a team of AUVs to update the map used by these vessels and, with it, perform mission-specific goals.
And in \cite{SVGP_Bart}, a sparse and variational Gaussian process is trained with datasets of different seafloor textures and then used for seafloor texture classification.
Our approach, in contrast, uses satellite data as prior in a Gaussian Process regression framework that is similarly updated with local measurements from AUVs to create high-resolution estimates of the environment around the autonomous agents.

\subsection{Contribution}

The main contributions of this paper are an algorithm containing a control law and a GP estimator, a sensitivity analysis with estimation method comparison, and an experimental setup and experimental demonstration in the Baltic~sea. 
In greater detail, this includes:

We implement a path-planning guidance law for adaptive sampling for the AUV heading. 
This control law receives information on the position of the AUV, the {\Chl} concentration measurement it took, and the gradient of the {\Chl} concentration gradient.
We analyze the performance of the control law for a {\Chl} concentration sensor uncertainty of $10^{-3}\text{mg/m}^3$, with full AUV and {\Chl} concentration sensor model.
In this scenario, the front tracking error concerning the {\Chl} concentration front reference is always below $0.02$~mg/$\text{m}^3$.

We propose a gradient estimator using GPs that estimates the gradient of {\Chl} concentration.
As it moves, the AUV records its position and {\Chl} concentration. 
The estimation is updated at each timestep by fitting a GP model with the latest data points collected by the AUV.
We analyze the performance of the GP gradient estimator for a {\Chl} concentration sensor uncertainty of $10^{-3}\text{mg/m}^3$, with full AUV and {\Chl} concentration sensor model.
In this scenario, the gradient estimation error concerning the {\Chl} concentration gradient field is always below $0.4$~rad, corresponding to less than $22$ degrees.

We design an experimental setup consisting of a cyber-physical system integrating the AUV software, the AUV hardware, the user interface, and a realistic simulator.
The AUV software includes numerous packages that can be divided into the behavior tree, the algal bloom front tracking, the onboard controllers, and the dead-reckoning.
The algal bloom front tracking library has been developed for the present work and includes the control law and two implemented estimation methods. 
It is available as an open-source contribution in \url{https://github.com/JoanaFonsec/gp4aes}.
The algal bloom front tracking package was also developed for the present work and is responsible for interfacing with the other AUV packages, simulating the {\Chl} sensor, and plotting.
It is available as an open-source contribution in \url{https://github.com/JoanaFonsec/algalbloom-tracking}.

We run a sensitivity analysis in which we vary the standard deviation of the {\Chl} sensor and measure the impact on the performance of the control law and the estimation method, considering the full AUV and {\Chl} sensor model. 
The {\Chl} sensor uncertainty we consider is between $10^{-3}\text{mg/m}^3$ and $10^{-1}\text{mg/m}^3$. 
This is in line with the {\Chl} concentration sensors in the market, which have a resolution of $10^{-2}\text{mg/m}^3$.
For this analysis, we consider two estimation methods, one based on Gaussian Process regression and the other based on Least Squares regression.
The analysis confirms that the control law tracking error and the gradient estimation error increase as the standard deviation of the {\Chl} sensor increases.
Particularly, for a standard deviation equal and above $10^{-2}\text{mg/m}^3$, the gradient estimation error increases exponentially for the LSQ estimator and linearly for the GP estimator.

We provide experimental results from two surveys in the Stockholm archipelago in the Baltic Sea. 
In these experiments, we demonstrate that the proposed algorithm performs well in the real-time real-world scenario and compare them to a simulation under experiment conditions.
Our results indicate that the front tracking error never exceeds 10 meters of overshoot and undershoot and that the gradient estimation error never exceeds 90 degrees.
We also examine the sources of error, namely surface waves that influence the AUV's movement but also partially occlude the GPS receiver, which introduces Gaussian noise on the GPS-measured position of the AUV.

\subsection{Organization}

The paper is organized as follows.
In Section~\ref{s: method}, we introduce the proposed front tracking algorithm. This includes the high-level system architecture, the dataset we use, the GP model for the chlorophyll-a concentration in the Baltic~sea, and the path planning guidance law.
An overview of the hardware and software components used in the AUV implementation is given in Section~\ref{s: implementation}.
In Section~\ref{s: simulations}, we provide realistic simulation results and a sensitivity analysis on the impact of sensor noise on algorithm performance with a comparison and evaluation of gradient estimation methods.
Experimental results follow in Section~\ref{s: experiments}, from field trials in the Baltic~sea in the summer of 2022.
Concluding remarks, discussion, and future directions follow in Section~\ref{s: conclusions}.

\section{Algal bloom front tracking}\label{s: method}

\begin{figure}[btp]
\centering
\includegraphics[width=.49\textwidth]{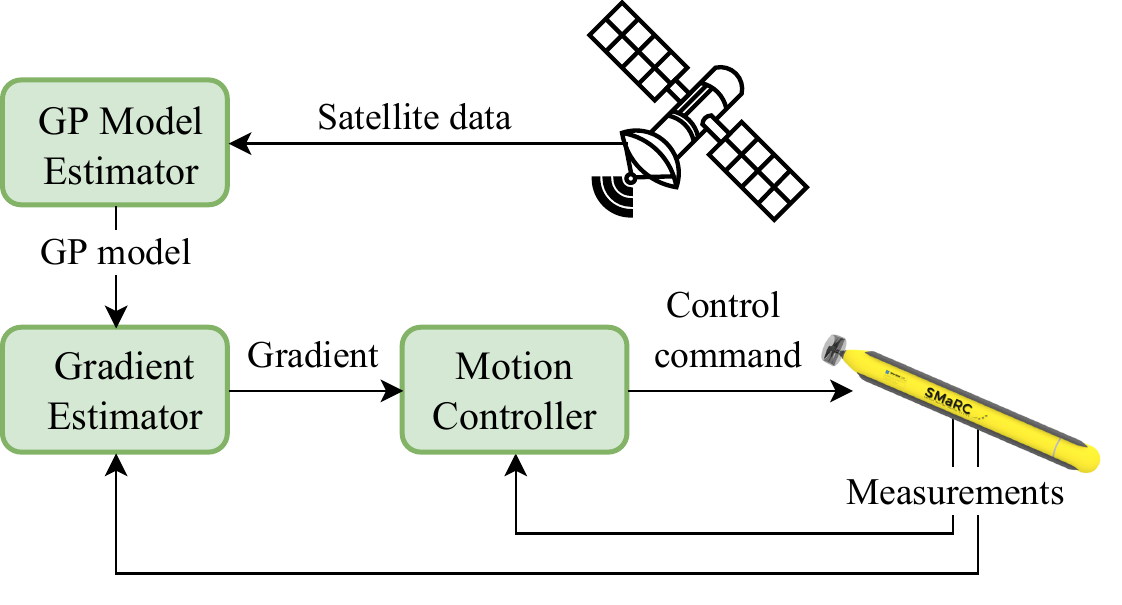}
\caption{System architecture with its main components: satellite data, GP model estimator, gradient estimator, motion controller, and AUV.} \label{system-architecture}
\vspace{-3mm}
\end{figure}

This paper considers algal bloom front tracking as an adaptive environmental sampling problem.
The objective of front tracking is to find and track a front with limited global information on its location or shape and only use local information collected by the AUV as it moves to explore the map.
This limited global information consists of satellite imagery from previous days to inform our model.
Then, the AUV has to decide where to explore next, given the information it has collected so far.
We approach this problem using an AUV with a {\Chl} sensor and remote satellite data from CMEMS.
Our solution consists of a novel system architecture containing three main components, as seen in the three green blocks of Fig.~\ref{system-architecture}.
They are a GP model estimator, a gradient estimator, and a motion controller.
In the following subsections, we present the system architecture and its components. 

\subsection{System Architecture}\label{sub: system}

The system architecture is summarised in Fig.~\ref{system-architecture}.
Here, we illustrate the main components of the proposed system, from the AUV to the motion controller, gradient estimator, GP model estimator, and satellite data.

The AUV has a {\Chl} sensor that measures the {\Chl} concentration at a set frequency as it moves in the field. The AUV movement is dictated by the control command received from the motion controller.
The motion controller uses the AUV's past measurements and an estimation of the field gradient to calculate the control command, closing the adaptive sampling loop.
The gradient estimator uses the past measurements taken by the AUV and a model of the {\Chl} concentration field to estimate the {\Chl} concentration field gradient.
The GP model estimator uses the previous days of satellite data to train kernel parameters of a GP model that represents the field we want to estimate. 
Finally, the satellite data consists of remote measurements of the {\Chl} concentration field from a few days preceding the mission and is used in the GP model estimator to generate the GP model estimate of the {\Chl} concentration field.

\subsection{Satellite Data}\label{sub: dataset}

\begin{figure} [btp] 
\vspace{1mm}
    \centering
    \includegraphics[width=.48\textwidth,trim={4cm 16cm 4cm 8cm},clip]{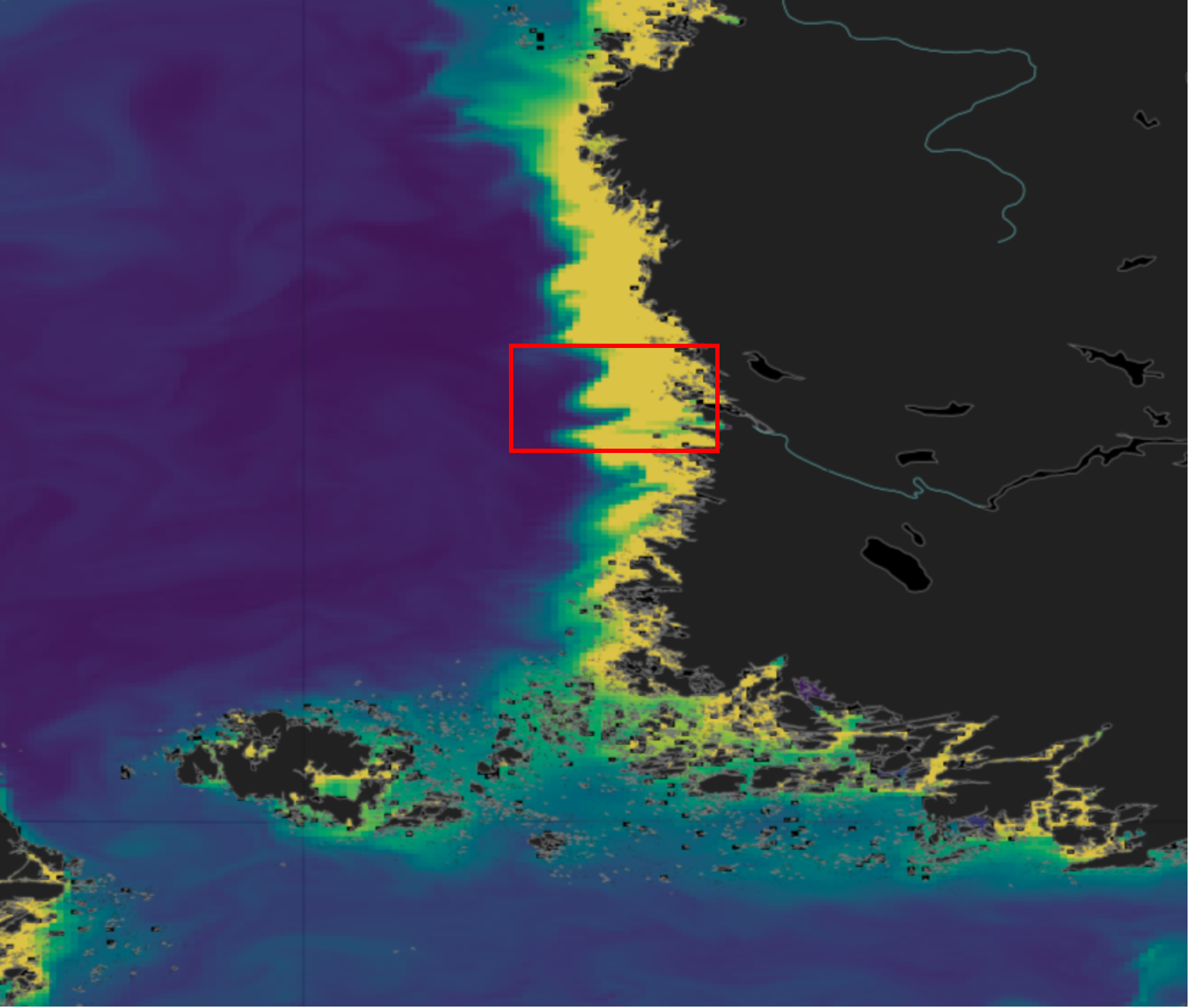} 
    \caption{CMEMS data of {\Chl} in the Baltic Sea (blue-yellow) and land (dark grey).}
    \label{fig:low_res}
    \vspace{-5mm}
\end{figure}

The satellite data concerns {\Chl} concentration for a given region.
We denote this {\Chl} concentration field by $\delta(\mathbf{p})$, where $\delta$ denotes the {\Chl} concentration at position $\mathbf{p}$.
In this paper, we consider surface data only.
Fig.~\ref{fig:low_res} shows a plot of sample {\Chl} concentration data, where high regions of high concentration are highlighted in yellow and regions of low concentration are highlighted in blue.
The dark grey area represents the land.
This data has a spatial resolution of 2~km by 2~km and is obtained from CMEMS \cite{CMEMS2km}.
The location is on the west coast of Finland, near the coastal city \textit{Pori}. 
We chose this location because a clear {\Chl} bloom front can be observed here. 
We hypothesize this is due to the nutrients that the river \textit{Kokem{\"a}enjoki} carries into the Baltic sea \cite{Pori}.
This paper will focus on the region marked by the red square taken on the 17th of April 2021. 
We chose April because it is the spring season of algal blooms.

\subsection{GP Model Estimator}\label{sub: GP}

The GP model estimator models the {\Chl} concentration field $\bar{\delta}(\mathbf{p})$ for the given region and time, exploiting prior information from satellite data of the previous days and measurements taken by the AUV in real-time.
Given this application's slow time scale, we assume that the {\Chl} concentration fields at different days have identical distributions, which simplifies model fitting and reduces the computational complexity. 

To obtain the {\Chl} concentration model, we must define the type of kernel that will accurately depict the process.
The kernel represents \textit{a priori} knowledge about the process by specifying how the {\Chl} concentration data is related to the corresponding spatial locations. 
Among the multitude of kernels described in the literature, only some fit the characteristics of the biogeochemical data we consider.
We use the Matérn kernel~\cite{Rasmussen}, which proves to be capable of modeling different degrees of smoothness across both vertical and horizontal length scales \cite{erickson2015gaussian}.
The covariance matrix $K\in\mathbb{R}^{N\times N}$ is defined concerning two points in the field map, $x_i$ and $x_j$.
Each matrix element is computed by the kernel function $k(x_i, x_j)$ for which $1 \leq i,j \leq N$. 
The kernel is defined as
\begin{equation} \label{eq:Kernel}
   K_{i,j} = k(x_i, x_j) = \sigma_k^2 (1 + r) e^{-r},
\end{equation}
where $r^2 = (x_i - x_j)^{\top} M (x_i - x_j)$, with
\begin{equation} \label{matrixM}
    M = \begin{bmatrix}
            \left(\frac{\sqrt 3}{l_0}\right)^2 & 0\\
            0 & \left(\frac{\sqrt 3}{l_1}\right)^2
        \end{bmatrix}.
\end{equation}
The kernel hyper-parameters are $(\sigma_k^2, l_0, l_1)$, where $\sigma_k^2$ is the variance of the {\Chl} concentration process, and $(l_0, l_1)$ define the length scale in each dimension.

The kernel hyper-parameters, $(\sigma_k^2, l_0, l_1)$, are estimated by maximizing the log marginal likelihood of the prior distribution - using only the available satellite data from previous days.
This data is called the training set and consists of a vector of size $N$ containing positions in the {\Chl} concentration field $\mathbf{X}=[\mathbf{p}_1,...,\mathbf{p}_N]$, and their respective chlorophyll~\textit{a} concentration values $\mathbf{y}=[\delta_1,...,\delta_N]$.
The log marginal likelihood to maximize is
\begin{align} \label{eq:log_marginal_likelihood}
    &\text{log}~p(\mathbf{y}|\mathbf{X}) = \\\nonumber &-\frac{1}{2}\mathbf{y}^{\top}(K+\sigma^2I)^{-1}\mathbf{y} - \frac{1}{2}~\text{log}~|K+\sigma^2I| - \frac{N}{2}\text{log}~2\pi
\end{align}
where $K$ is the $N\times N$ covariance matrix in which each value is created as in \eqref{eq:Kernel}, and $\sigma$ is the noise variance of each data point.

Using the trained kernel, the GP model can be fitted with the collected data, using the standard conditioning formulae~\cite{Rasmussen} to obtain the model for the {\Chl} concentration field, which we define as $\bar{\delta}(\mathbf{p})$.
To do so, we consider the $n$ most recent measurements taken by the AUV. 
It contains the AUV's positions $\mathbf{P}=[\mathbf{p}_1,...,\mathbf{p}_n]$, and its chlorophyll~\textit{a} concentration measurements $\mathbf{\Delta}=[\delta_1,...,\delta_n]$.
Then, the mean of the model $\delta(\mathbf{p})$ is denoted by $\bar{\delta}(\mathbf{p})$ and the covariance by $\text{cov}{\left(\delta(\mathbf{p})\right)}$.
We can compute the mean and covariance at some point $\Position_*$ as
\begin{align} \label{eq:gp}
    \bar{\delta}(\mathbf{p}_*) &= K_* \left(K + \sigma^2 I \right)^{-1}\mathbf{\Delta}\\ \label{eq:cov}
    \text{cov}{\left(\delta(\mathbf{p}_*)\right)} &= K_{**} - K_* \left[K + \sigma^2 I\right]^{-1}K_*^T
\end{align}

where $K\in\mathbb{R}^{n\times n}$ corresponds to the covariance between the data in points $\mathbf{P}$, $K_*\in\mathbb{R}^{1\times n}$ corresponds to the covariance between the data in points $\Position_*$ and $\mathbf{P}$, $K_{**}=\sigma^2$ corresponds to the variance at the point $\Position_*$, and $\sigma^2$ is the variance of the measurement noise.
Since the set of the $n$ most recent measurements is changing in time, then our estimate $\bar{\delta}(\mathbf{p}_*)$ is also changing in time.

\medskip
To evaluate the accuracy of the GP model applied to {\Chl} concentration fields, the kernel is trained using both the low and high-resolution datasets.
Then, the goodness of fit of both sets of parameters is evaluated by comparing the respective predictions of the {\Chl} concentration to the ground truth.
Using past satellite data, we construct the set $X$ and $y$ in \eqref{eq:log_marginal_likelihood} from data from multiple process realizations.
Such division of the training dataset prevents overfitting.
Then, $X$ and $y$ are composed of data from 3 days before the prediction date, selecting non-overlapping random scattered sub-datasets of the same size each day. 
The optimization algorithm for the maximization of \eqref{eq:log_marginal_likelihood} is L-BFGS-B, which is implemented in the \texttt{scipy} library, and the resulting parameters follow in Table~\ref{tab:params}.

\begin{table}[hbtp]
    \centering
    \begin{tabular}{c|c|c|c}
         & $\sigma^2$   & $l_0$  & $l_1$ \\ \hline
        Low res & 44.2959 & 0.5465 & 0.2890 \\ \hline
        High res & 18.2106 & 0.0559 & 0.0245
    \end{tabular}
    \caption{Kernel hyperparameters obtained through maximum likelihood estimation, using the low-resolution and high-resolution datasets.}
    \label{tab:params}
\end{table}

\begin{figure} [btp]
\vspace{1mm}
\centering
  \begin{subfigure}[b]{\linewidth}
    \includegraphics[width=\linewidth,trim={2mm 18mm 0mm 29mm},clip]{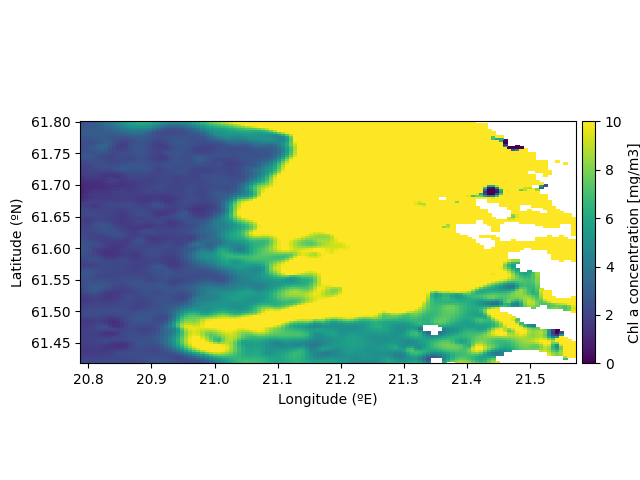}
    \vspace{-6mm}
    \caption{Predicted mean using lower resolution data.}      \label{fig:scattered_low_res}
  \end{subfigure}\\
  \begin{subfigure}[b]{\linewidth}
    \includegraphics[width=\linewidth,trim={2mm 18mm 0mm 29mm},clip]{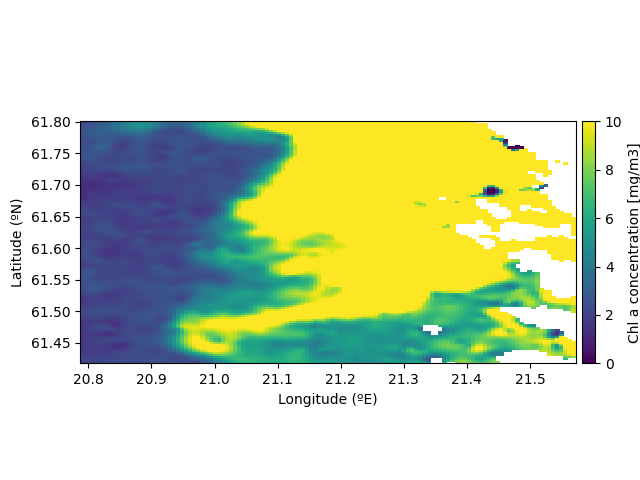}
    \vspace{-6mm}
    \caption{Predicted mean using higher resolution data.}
    \label{fig:scattered_high_res}
  \end{subfigure}\\
  \begin{subfigure}[b]{\linewidth}
    \centering
    \includegraphics[width=\linewidth,trim={3mm 18mm 0mm 29mm},clip]{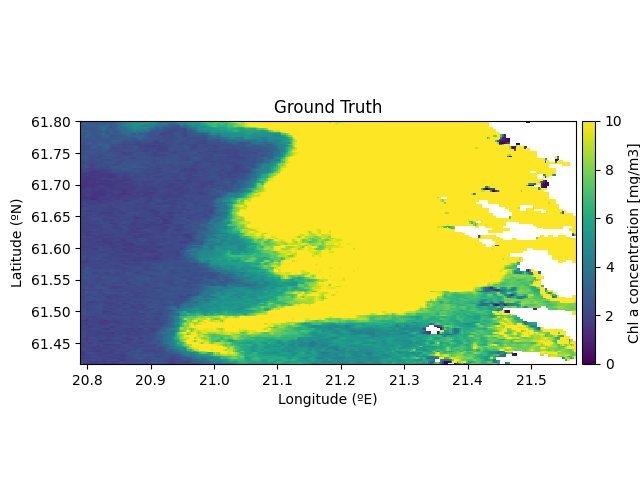}
    \vspace{-6mm}
    \caption{Higher resolution dataset that simulates the ground truth.} 
    \label{fig:high_res_real}
  \end{subfigure}
    \caption{Predicted mean from scattered measurements compared to the ground truth (high-resolution data).}
\vspace{-5mm}
\end{figure}

The results of this comparison are presented in Figures~\ref{fig:scattered_low_res} and \ref{fig:scattered_high_res}. 
The observations and test datasets are approximately of size 1500 and 13500, respectively, having in consideration a ratio of 10/90\% between both, where the former is a set of scattered samples from the ground truth data in figure \ref{fig:high_res_real}, having a standard deviation of $\sigma_n = 10^{-3}$. 

Visually, the results are very similar. 
The average relative error of the prediction compared to the ground truth data in figure \ref{fig:high_res_real} was approximately 12\% and 11\% using low and high-resolution data, respectively.
Based on the similarity between results, we conclude that the proposed GP model accurately represents the statistical properties of the chlorophyll \textit{a} concentration in the operations scenario, even when the training dataset is different from the ground truth data.

\subsection{Gradient Estimator}

The gradient estimator derives the previously obtained model of the {\Chl} concentration to estimate the {\Chl} concentration gradient field.
From the equation \eqref{eq:gp}, the gradient $\nabla\bar{\delta}(\mathbf{p}_*)$ is obtained by computing the derivative of the predicted {\Chl} concentration with respect to position $\Position_*$,
\begin{equation}
\label{eq:pred_mean_grad}
    \nabla\bar{\delta}({\Position_*}) = {\nabla_{\Position_*}}{\left[ K_* \left(K + \sigma^2 I \right)^{-1}\mathbf{\Delta} \right]}.
\end{equation}
Since the second and third terms inside the gradient in \eqref{eq:pred_mean_grad} are constant relative to ${\Position_*}$, we only need to compute $\nabla_{{\Position_*}}K_*$. 
Each element of the $K_*$ matrix is given by \eqref{eq:Kernel}, in which $x_i$ corresponds to $\Position_*$ and $x_j$ corresponds to $\mathbf{p}_j \in \mathbf{P}$.
So we take the derivative of $k(\Position_*, \mathbf{p}_j)$ with respect to $\Position_*$,
\begin{equation*}
    \nabla_{\Position_*} k(\Position_*, \mathbf{p}_i) = -\sigma_k^2 e^{-r} M (\mathbf{p}_* - \mathbf{p}_j),
\end{equation*}
where $M$ and $r$ are as in subsection~\ref{sub: GP}.
Note that the gradient of the kernel equation is not defined when the test point in $\mathbf{P}$ is equal to the current position $\Position_*$. 
To account for this, the current position $\Position_*$ is not included in $\mathbf{P}$ when computing \eqref{eq:pred_mean_grad}.
Then the gradient estimate at position $\Position_*$ is
\begin{equation}
\label{eq:gradient}
    \nabla\bar{\delta}({\Position_*}) =  \nabla_{\Position_*} K_* \left(K + \sigma^2 I \right)^{-1}\mathbf{\Delta}.  
\end{equation}

\subsection{AUV}\label{sub: AUV}

The AUV receives the control command $\mathbf{u}$ from the motion controller, which is the reference for direction and velocity.
Then, using its internal lower-level controller, the AUV turns this reference $\mathbf{u}$ into thrust commands to its thrusters $\boldsymbol{\tau_C}$.
For this AUV, we consider the 6DOF model in which the state is the velocity vector given by $\boldsymbol{\nu}=\left[\begin{array}{cccccc} u & v & w & p & q & r \end{array}\right]^T $ containing the translational and rotational velocities.
Note that these velocity vector elements directly influence the AUV's position $\Position$. 

The dynamics of the AUV can be formulated as a nonlinear system represented by a vectorial notation presented by Fossen \cite{Fossen:2011} as follows:
\begin{multline}
    \label{eq:Fossen Flight Dynamics}
    \boldsymbol{(M}_{RB} + \boldsymbol{M}_A)\dot\nu + \boldsymbol{(C}_{RB}(\nu) + \boldsymbol{C}_{A}(\nu))\nu +  \\
   \boldsymbol{ D}(\nu)\nu + \boldsymbol{g}(\eta) = \boldsymbol{\tau_C} \; ,
\end{multline}
where $\boldsymbol{M}_{RB}$ is the rigid body mass and inertia matrix and $\boldsymbol{C}_{RB}$ is the matrix of Coriolis and centripetal terms on the left-hand side. 
$\boldsymbol{M}_A$ and $\boldsymbol{C}_A(\nu)$ represent the effect of added mass, $\boldsymbol{D}(\nu)$ represents the damping matrix, and $\boldsymbol{g}(\eta)$ is the vector of gravitational and buoyancy forces and moments.
$\boldsymbol{\tau_C}$ is a vector of external control forces based on the AUV's actuator configuration. The damping matrix $\boldsymbol{D(\nu)}$, in particular, has a significant effect on the nonlinear hydrodynamics of the AUV \cite{BhatStenius2020} and is a key simulation parameter. 

\subsection{Motion Controller}\label{sub: guidance}

\begin{figure}[btp]
\vspace{2mm}
    \centering   \includegraphics[width=\columnwidth,trim={0 2mm 2mm 0}, clip]{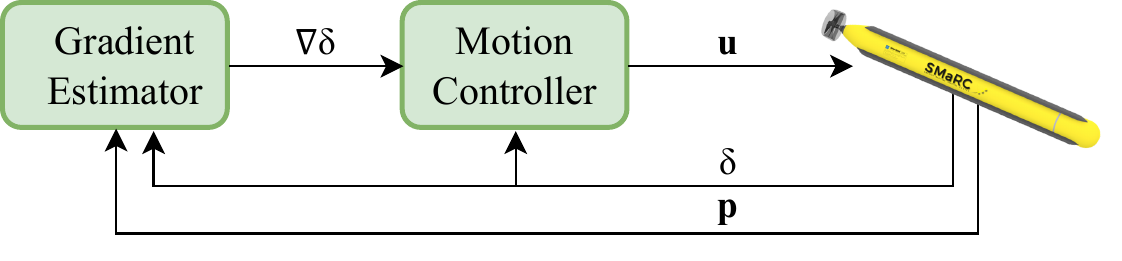}
    \caption{Control architecture with the motion controller, the gradient estimator, and the AUV.}
    \label{control_archi}
    \vspace{-5mm}
\end{figure}

The control law we propose is summarised in Fig.~\ref{control_archi}.
It relies on the {\Chl} gradient $\nabla\delta$ and the latest {\Chl} concentration measurement $\delta$ to produce a control command $\mathbf{u}$.
First, we define a front as a level set of a time-varying scalar field $\delta : \mathbb{R} \times \mathbb{R}^2 \to \mathbb{R}$:
\begin{equation}\label{eqn:front}
    F(t) = \{ \mathbf{p} \in \mathbb{R}^2 : \delta(t, \mathbf{p}) = \delta_\mathrm{ref} \},
\end{equation}
where $\delta_\mathrm{ref}$ is some reference value, $\mathbf{p}$ the position and $t$ time.

Assuming that the reference value $\DeltaRef$ is known, we used the previously developed control law as in \cite{joana2021}. 
There, we define the control law as

\begin{equation}\label{eqn:law}
\begin{aligned}
    \mathbf{u}(t, \Position) &= \mathbf{u}_{\text{seek}}(t, \mathbf{p})
        + \mathbf{u}_{\text{follow}}(t, \Position) \\
    \mathbf{u}_{\text{seek}}(t, \Position) &= -\alpha_{\text{seek}}(\delta(t, \mathbf{p}) - \DeltaRef)\nabla\delta(t, \mathbf{p}) \\
    \mathbf{u}_{\text{follow}}(t, \Position) &= \alpha_{\text{follow}} R_{\pi / 2}\nabla\delta(t, \mathbf{p}),
\end{aligned}
\end{equation}

where $\nabla\delta$ is the gradient of $\delta$ with respect to $\Position$, $R_{\pi / 2}$ is a mapping which rotates vectors by 90~degrees, and $\alpha_{\text{seek}}$ and $\alpha_{\text{follow}}$ are tunable parameters.

\begin{figure}[btp]
    \centering
    \includegraphics[width=\columnwidth,trim={2mm 4mm 2mm 11mm}, clip]{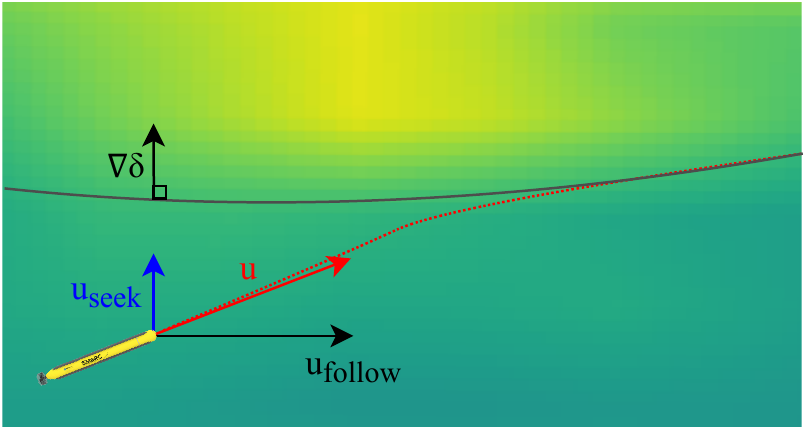}
    \caption{Seek and follow components of the control law and gradient.}
    \label{Control}
    \vspace{-5mm}
\end{figure}

As seen in Fig.~\ref{Control}, the control law consists of two components: $\mathbf{u}_{\text{seek}}$, which controls the AUV towards the front by following the gradient field, and $\mathbf{u}_{\text{follow}}$ which controls the AUV to move along the front, perpendicularly to the gradient field.
By designing the control law with these two components, we ensure convergence to the front \cite{joanacontrolo}.
Namely, note that if $\delta(t,\mathbf{p}) \ne \DeltaRef$, the $\mathbf{u}_{\text{seek}}$ component grows proportionally to this difference, making seeking the front a priority, comparing to following the front.
On the other hand, if the AUV is on the front, the most prominent component becomes the $\mathbf{u}_{\text{follow}}$.


\section{Experimental Setup}\label{s: implementation}

\begin{figure*}[htb]
    \centering
    \includegraphics[width=\textwidth,trim={0 0 0 0}, clip]{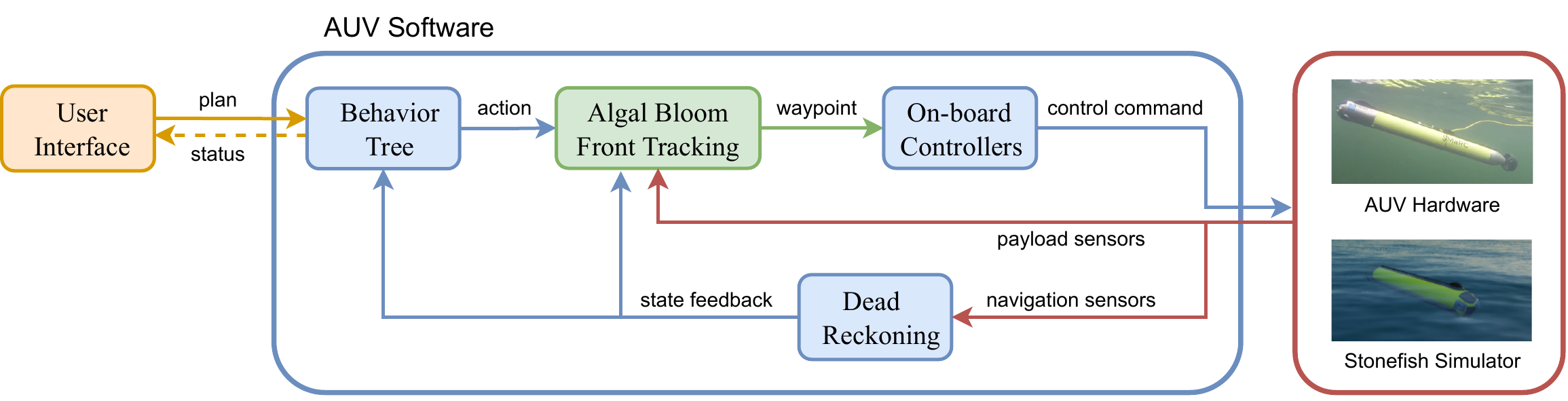}
    \caption{The cyber-physical system architecture integrates the user interface, AUV software, AUV hardware, and simulator.}
    \label{fig:sam_sys_diag}
    \vspace{-3mm}
\end{figure*}

The test platform used in these experiments is the SAM AUV, a research platform developed at the Swedish Maritime Robotics Centre (SMaRC) \footnote{\url{https://www.smarc.se}}. The algal bloom front tracking algorithm from the previous section is integrated into the SAM AUV software to be deployed in the field and validated experimentally. 

The experimental setup is a cyber-physical system (CPS) whose architecture is summarised in Fig. \ref{fig:sam_sys_diag}. The user interacts with the AUV software system through an interface that enables them to send mission plans and monitor the current status. A behavior tree monitors the mission status and delegates actions to an algal bloom front tracker. The algal bloom front tracker reads payload measurements of {\Chl} data and sends waypoints to onboard feedback controllers. These onboard controllers, in turn, send commands to actuators considering state feedback. Such state feedback is provided by dead-reckoning based on onboard navigation sensors on the AUV hardware. The entire software system can also be validated in simulation via the Stonefish simulator \cite{stonefish}. Each component of the CPS will be further described in the following subsections.

\subsection{AUV Hardware}\label{sub: SAM hardware}

\begin{figure}[btp]
    \centering
    \includegraphics[width=\columnwidth,trim={3.7cm 3.5cm 4.7cm 4cm}, clip]{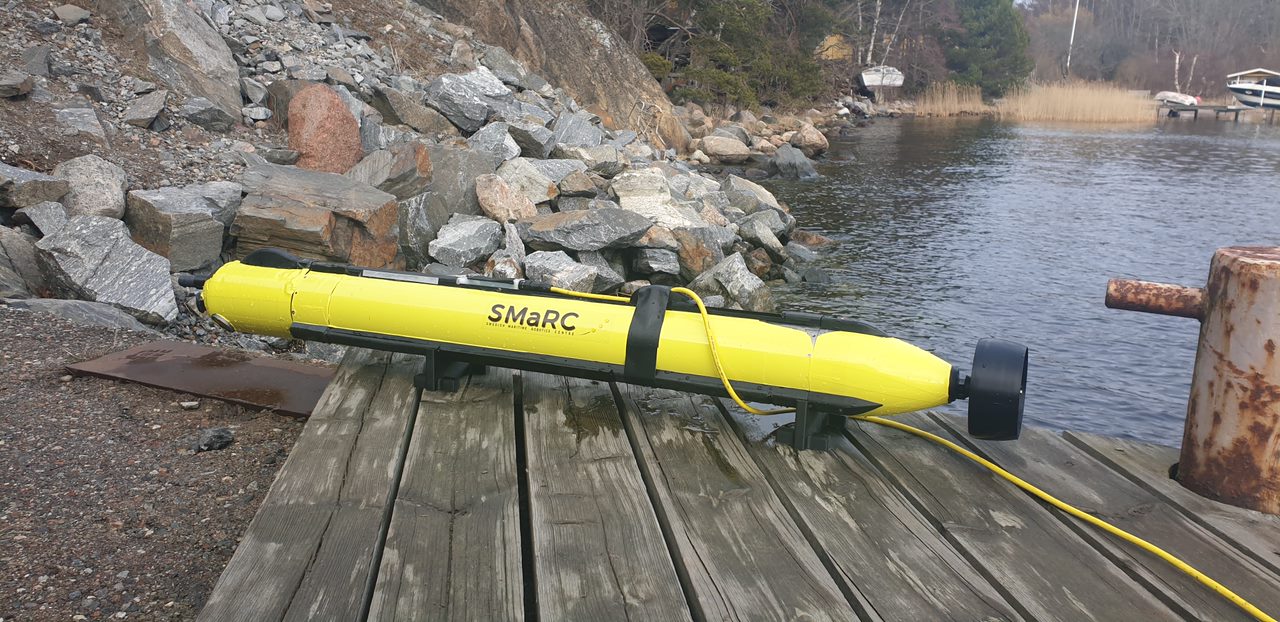}
    \caption{The SAM AUV developed by SMaRC.}
    \label{SAM_pic}
    \vspace{-3mm}
\end{figure}

We begin by describing the SAM AUV hardware in the upper right corner of Fig.~\ref{fig:sam_sys_diag}. 
SAM, short for Small and Affordable Maritime robot (as seen in Fig.~\ref{SAM_pic}), is a torpedo-shaped, under-actuated AUV that serves as an agile research platform for testing new sensing, perception, and control strategies \cite{Bhat_etal_2019_CPS, Bhat_etal:2020b}.  

\begin{figure}[btp]
    \centering
    \includegraphics[width=\columnwidth,trim={0 0 0 7mm}, clip]{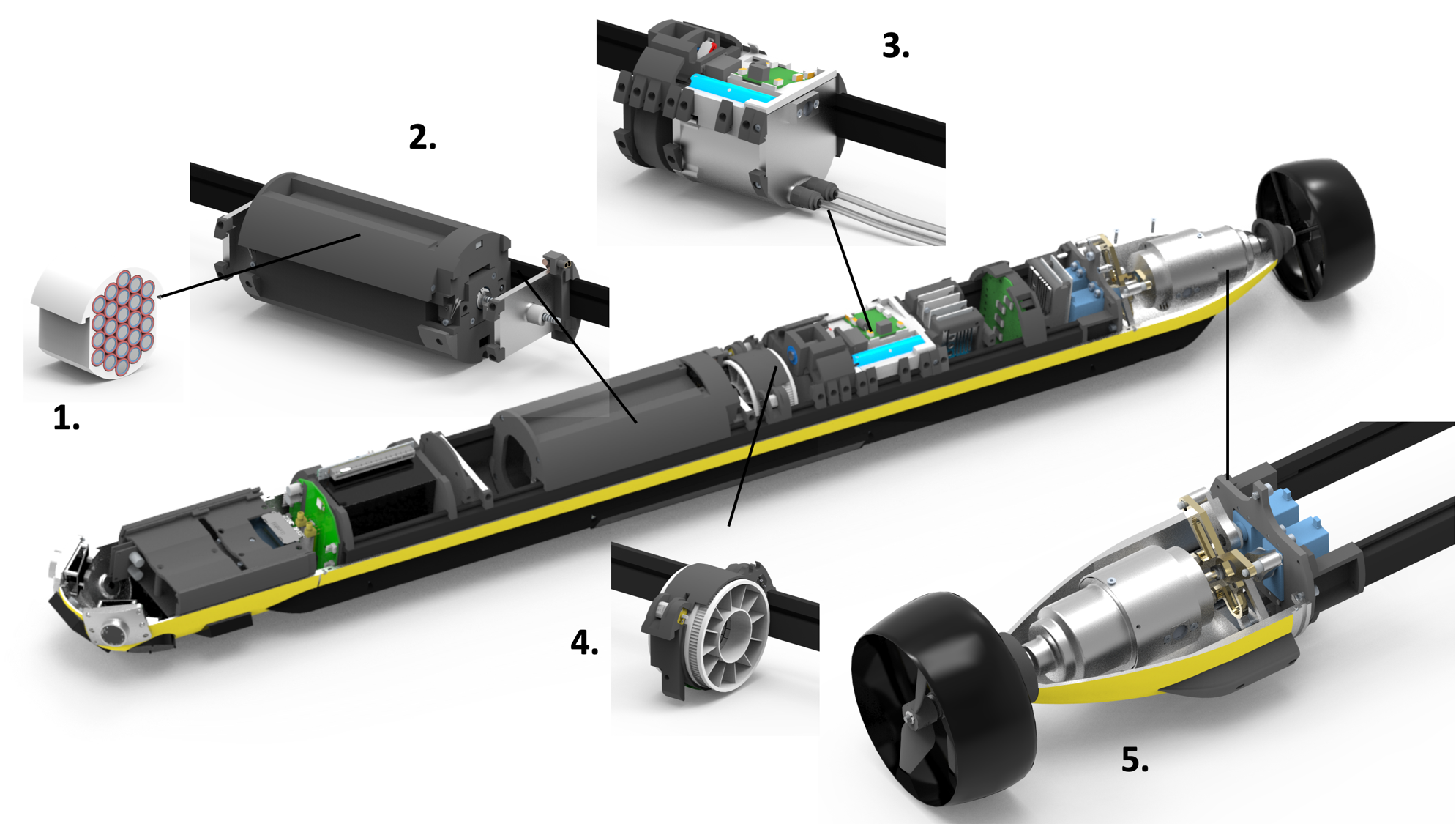}
    \caption{AUV subsystems: 1. Battery pack, 2. Longitudinal Center of Gravity Trim System, 3. Variable Buoyancy System, 4. Transversal Center of Gravity System, 5. Thrust Vectoring System with Counter-rotating Propellers. }
    \label{fig:HardwareSystems}
    \vspace{-5mm}
\end{figure}

 Five key actuator subsystems enable SAM to be highly maneuverable and hydrobatic (these are depicted in Fig. \ref{fig:HardwareSystems}):  
 \begin{enumerate}
     \item The \textit{Longitudinal Center of Gravity (LCG)} system uses a movable battery pack to change the center of gravity position longitudinally and enable static pitch control.
     \item The \textit{Transversal Center of Gravity (TCG)} system contains rotating counterweights that enable static roll control or changes to the AUV's stability margin.
     \item The \textit{Variable Buoyancy System (VBS)} facilitates buoyancy regulation and static depth control by pumping water in and out of a tank. 
     \item \textit{Counter-rotating propellers} provide propulsion while compensating for propeller-induced roll.
     \item The \textit{Thrust Vectoring} system contains a servo-actuated nozzle for steering in the horizontal and vertical planes. The thrust vectoring nozzle enables rapid maneuvering in the horizontal and vertical planes.
 \end{enumerate}

Sensors have been integrated into SAM for navigation and environmental sensing. Navigation sensors include an Inertia Measurement Unit, a compass, a GPS receiver, a Doppler Velocity Logger (DVL) for bottom tracking, and pressure sensors for depth measurements. Payload sensors include cameras and sidescan sonar for inspection and surveying, a forward-looking sonar for obstacle avoidance, and a Conductivity-Temperature-Depth (CTD) probe for water-column monitoring.
For our algal bloom front tracking application, we use a \Chl-turbidity-phycocyanin fluorometer for phytoplankton sensing.

\subsection{Stonefish Simulator}\label{sub: simulator}

Before running autonomy software on the SAM AUV, simulations of mission scenarios can be performed within the Stonefish simulator, represented in the bottom right corner of Fig.~\ref{fig:sam_sys_diag}.
Stonefish offers a photo-realistic simulation environment where entire missions can be modeled and rehearsed. The AUV's dynamics are modeled within the simulator together with models of the sensors. Objects, environmental features, and bathymetry can be imported into the simulator to create mockups of planned environments. Within Stonefish, perception and planning software can be validated before deployment on the hardware.  

For the application presented in this paper, the satellite data for algal blooms is modeled in Stonefish as a lookup table of {\Chl} values over a grid encompassing the entire simulation environment. A simulated chlorophyll sampler interpolates the relevant chlorophyll measurement from this grid based on the AUV's current position. The software interfaces within the Stonefish simulator and the real SAM AUV are identical, thus enabling virtual validation of a full mission sequence for algal bloom tracking.

\begin{figure}[btp]
    \centering
    \includegraphics[width=\columnwidth]{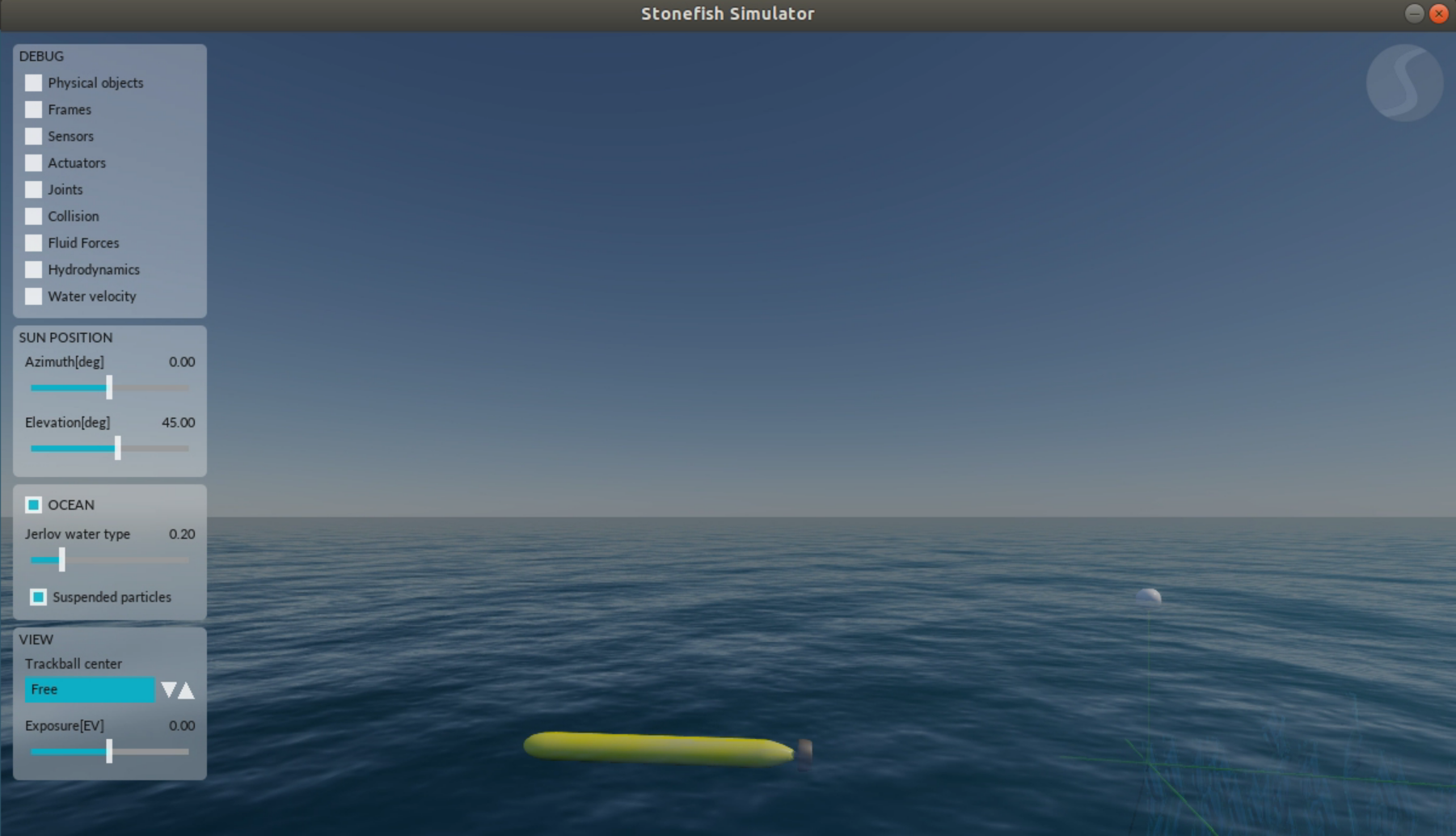}
    \caption{Print screen of the Stonefish simulator.}
    \label{fig: stonefish}
    \vspace{-5mm}
\end{figure}

\subsection{AUV Software}\label{sub: SAM software}

The four AUV software components presented in the center of Fig.~\ref{fig:sam_sys_diag} are further detailed in this section. 
The autonomy software of SAM runs on the Robot Operating System (ROS)\footnote{\url{https://www.ros.org}} environment. Its sub-components include a \textit{behavior tree} for decision-making and mission execution, a path planner for \textit{algal bloom front tracking}, \textit{onboard controllers} for path following, and a \textit{dead-reckoning} package for navigation. These software packages are also available in an open-source repository in \url{https://github.com/smarc-project}. 

\medskip
\subsubsection{Behavior Tree} 
SAM uses a Behavior Tree (BT) to ensure safe and transparent mission execution.  A BT is a reactive decision-making structure that is comprised of Sequences, Fallbacks, Actions, and Conditions. 
The main objective of the BT is to receive a mission plan and delegate actions (e.g., waypoints) to lower-level systems (e.g., motion planners and controllers) while checking the safety criteria of the AUV simultaneously and continuously. In an unsafe situation, the BT executes emergency actions to bring the system back to a safe state. A BT ensures safety and compliance requirements during mission execution by disallowing unsafe behaviors autonomously. Further information on designing BTs for underwater robots can be found in \cite{sprague2018improving, ozkahraman2020combining}.

\begin{figure}[btp]
    \centering
    \includegraphics[width=\columnwidth]{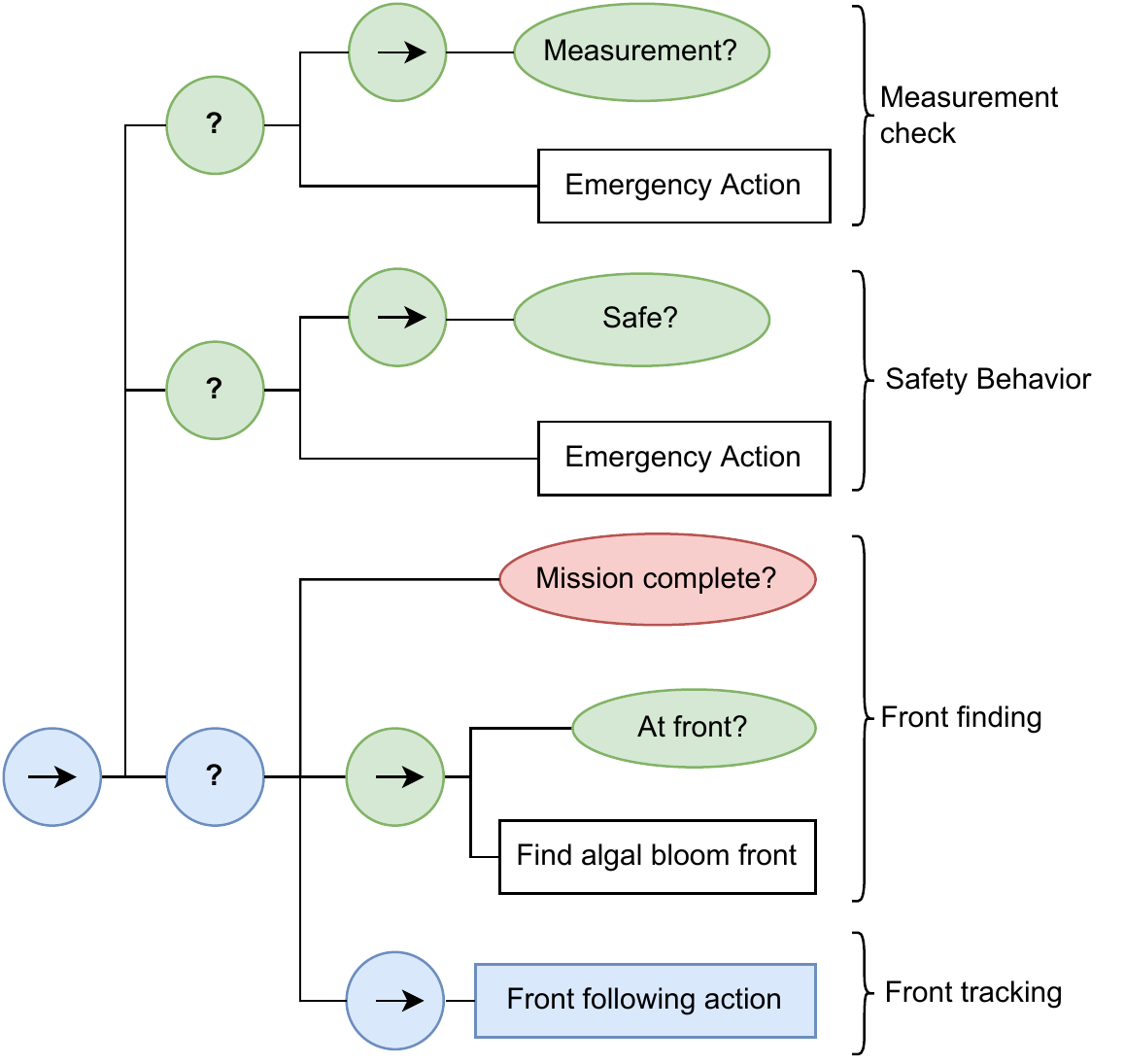}
    \caption{High-level view of the BT used for algal bloom front tracking. Inner nodes are sequences (arrows) and fallbacks (question marks). Leaf nodes are actions (rectangle) and conditions (ellipse). All nodes can return Success (green), Failure (red), and Running (blue).}
    \label{fig:BT}
    \vspace{-5mm}
\end{figure}

The Gaussian process path planner presented in Section \ref{sub: guidance} is integrated into the BT with additional conditions and actions considering satellite data ingestion, {\Chl} sampling, and front-tracking.  Each algal bloom tracking experiment using SAM follows a set operational sequence. Considering a relevant area with algal blooms, the following workflow is used for sampling and tracking the bloom: 
\begin{enumerate}
    \item An area of interest is specified, and satellite imagery is downloaded.
    \item An initial mission plan is set via a user interface for the AUV to traverse to the algal bloom feature.
    \item The AUV is launched, and a GPS fix is acquired.
    \item The AUV starts the mission and follows user-defined waypoints to reach the algal bloom front's vicinity.
    \item The AUV detects the front through a chlorophyll sensor, and the front-tracking algorithm is engaged.
    \item The path planner generates new waypoints to track the front upon sampling, and the AUV follows these waypoints to track the front.
\end{enumerate}
This sequence is also rehearsed within the Stonefish simulator.
To follow such a mission sequence, the BT for algal bloom tracking is summarised in Fig.~\ref{fig:BT}. 
A sanity check on chlorophyll measurements is performed in the first sub-tree. 
Second, safety conditions are checked. 
If either fails, the mission is aborted, and emergency actions are performed.  
Third, if the system is safe and measurements are available, the user-defined waypoint mission is followed. 
Fourth, if the AUV reaches the front, the algal bloom front following action is performed. This is further detailed below. 

Such a BT structure allows the designer to prioritize actions, sequence missions, and ensure the vehicle operates safely while keeping the overall setup modular and easy to understand.

\medskip
\subsubsection{Algal Bloom Front Tracking} 
The algorithm of the previous section is incorporated as the algal bloom front following action in the BT. The action ingests payload data on chlorophyll concentration and sends live waypoints to the onboard controllers so that the AUV follows the front.

When the AUV crosses the algae front, the front tracking behavior is enabled, with a higher priority than following the original waypoints (see Fig.~\ref{fig:BT}). A path planner for front tracking sends new waypoints to the AUV based on live measurements. The AUV samples the front and follows the edge of the bloom.  Once the AUV has exited the front, the vehicle will fall back to following the operator's plan. 

\medskip
\subsubsection{Onboard Controllers}

The waypoints sent from the original mission plan and the front-tracking algorithm provide input to the onboard controllers on SAM. These onboard controllers enable the AUV to follow pre-defined waypoints and track the algal front. Given a set of waypoints, a line-of-sight guidance law minimizes cross-track error and ensures the vehicle can approach each waypoint at a set heading and depth. Feedback controllers then control the AUV to follow these set points. Further information on the waypoint following guidance law can be found in \cite{Stenius2022}.

 In SAM's case, the control force vector $\boldsymbol{\tau_C(c)}$ in equation (\ref{eq:Fossen Flight Dynamics}) is a function of the available actuator inputs, which are contained in   
\begin{equation}
\boldsymbol{c}=
\left[ \begin{array}{*{6}{c}}
    rpm_1 & rpm_2 & d_e & d_r & LCG & VBS
\end{array}\right],
\label{eq-system-inputs-u}
\end{equation}
where $rpm_{1,2}$ represent the propeller speed, $d_e$ and $d_r$ are vertical and horizontal thrust vector angles, and $LCG$ and $VBS$ specify the position and buoyancy level, respectively.

\begin{figure}[btp]
    \centering
    \includegraphics[width=\columnwidth,trim={21mm 37mm 24mm 15mm}, clip]{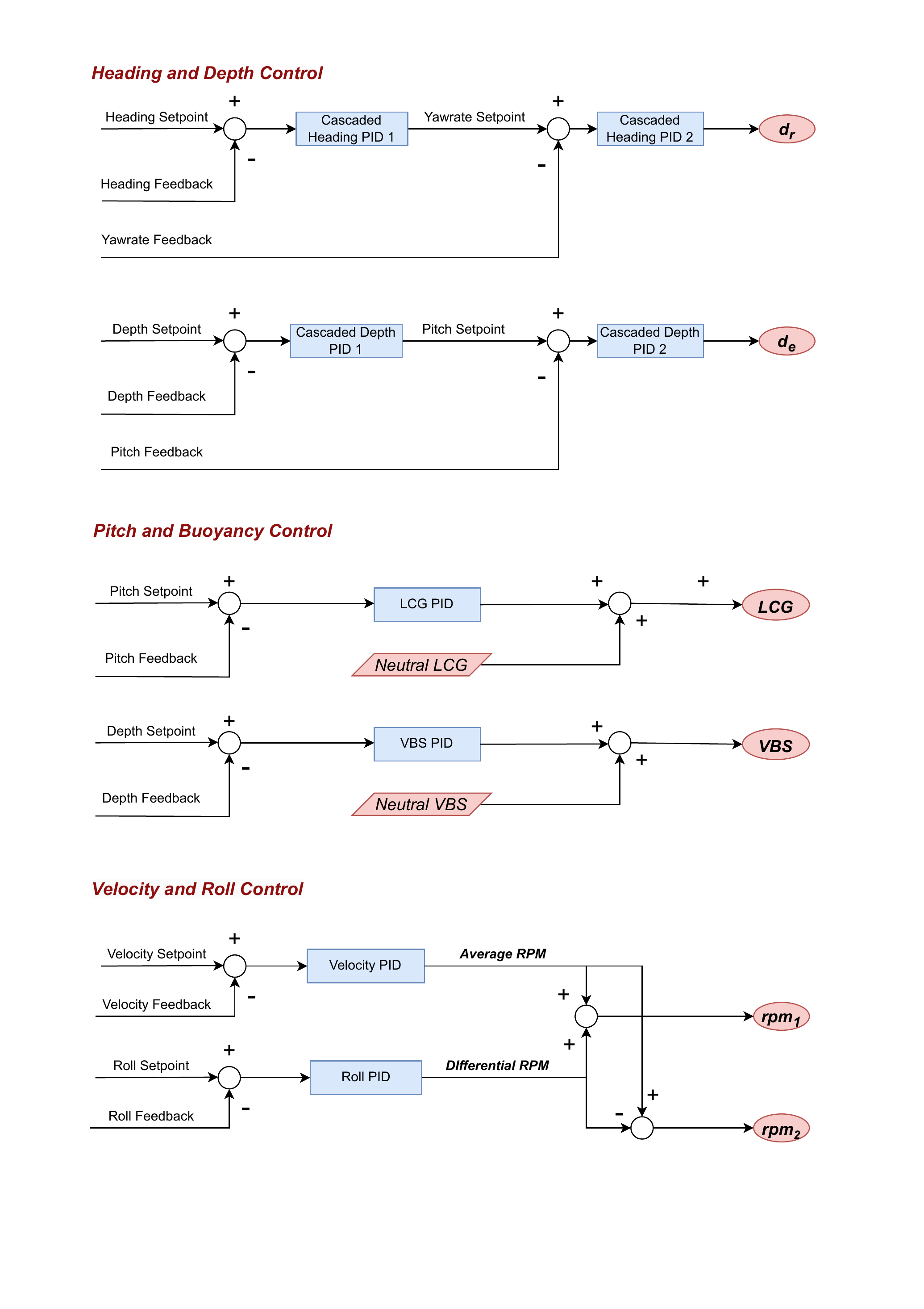}
    \caption{The feedback control architecture on SAM with cascaded heading and depth PIDs for flight control (top), trim stabilization with pitch and buoyancy control (middle), and coupled velocity and roll control with the counter-rotating propellers (bottom).}
    \label{fig: SAM Control}
    \vspace{-5mm}
\end{figure}

The feedback control architecture is presented in Fig.~\ref{fig: SAM Control}). First, we consider flight control to regulate the heading and depth where cascaded Proportional-Integral-Derivative (PID) controllers are used to command the thrust vector angles. In the outer loop, the controllers provide a yaw rate and pitch setpoint, which are translated to actuator commands to the thrust vectoring system in the inner loop. These account for couplings between states for flight control. Second, to stabilize the AUV in pitch and depth, additional PID controllers control the trim (LCG) and buoyancy (VBS) subsystems. Finally, coupled roll and velocity control are realized using parallel PIDs to command the counter-rotating propellers. These provide an average propeller rpm to achieve the desired velocity while also providing a differential rpm between the two propellers that causes the AUV to hold a roll angle. It is also possible to directly command constant rpm values instead of a desired velocity. The combination of flight and trim controllers enables SAM to track the algal front at a specified velocity or propeller rpm but are dependent on reliable state feedback.

\medskip
\subsubsection{Dead Reckoning} 
Underwater navigation is challenging because radio waves attenuate rapidly in water. 
This means that GPS-based positioning and navigation are not available underwater --- we need to use inertial and acoustic sensors for estimating the AUV's position, orientation, and velocities. 
Dead reckoning is thus used for obtaining state feedback.
An Extended Kalman Filter is used to fuse acoustic and inertial measurements collected by the AUV's onboard sensors to estimate the vehicle's current position, orientation, and velocity. 
In particular, the IMU and compass are used to obtain orientations, angular velocities, and accelerations, the DVL is used to obtain linear velocities, and the pressure sensor is used to calculate the depth.
These are then fused and integrated with a motion model to obtain a position and velocity estimate.

\begin{figure}[btp]
    \centering
    \includegraphics[width=\columnwidth,trim={5mm 68mm 11cm 0}, clip]{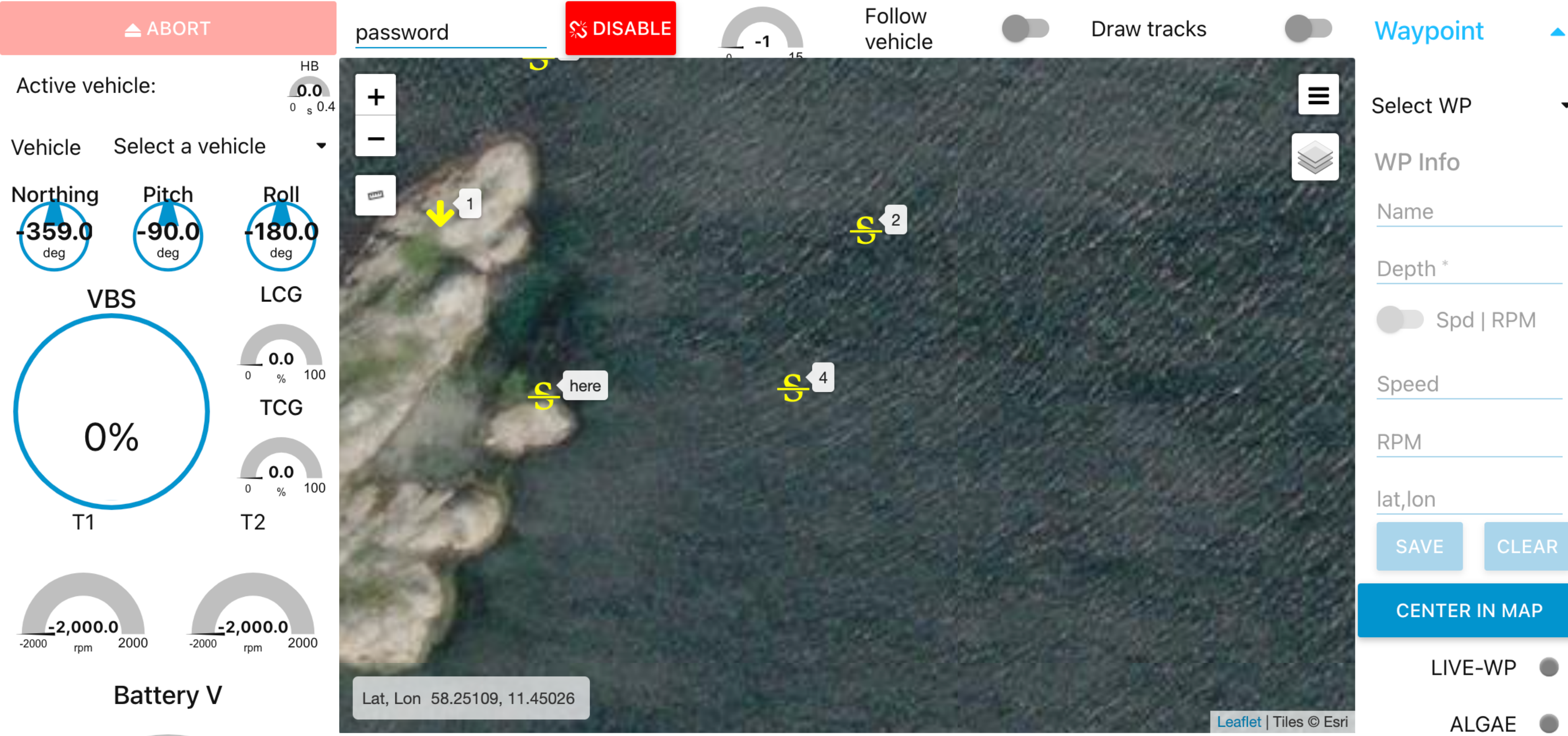}
    \caption{Print screen of the user interface: Node-Red.}
    \label{fig: interface}
    \vspace{-3mm}
\end{figure}

\subsection{User Interface}\label{sub: interface}

The user interacts with SAM's software system via the user interface in the left corner of Fig.~\ref{fig:sam_sys_diag}. 
As represented in Fig.~\ref{fig: interface}, it consists of a web-based graphical interface based on \textit{Node-RED}. It enables the operator to plan the mission on a world map and monitor the vehicle's current status during the mission. Different AUV parameters can be tracked, new missions can be run, and the current status of several measurements can be observed. A second graphical user interface allows test engineers to run specific hardware drivers and controllers on SAM for low-level control and validation.


\section{Simulation results}\label{s: simulations}

In this section, the control and estimation components of the proposed system architecture are tested in a front-tracking procedure in the operational area in  Fig.~\ref{fig:high_res_real}. 
The simulation starts by deploying the vehicle close to the front and providing an initial heading setpoint towards it.
When the AUV reaches the front, the gradient estimation is triggered, and the control law receives the estimated value from the GP model as an input.
The section is divided into three subsections: simulation setup, numerical results, method analysis, and comparison.

\subsection{Simulation setup}\label{sub: sim_setup}

\begin{figure} [btp] 
    \vspace{2mm}
    \centering    \includegraphics[width=.45\textwidth,trim={4cm 16cm 4cm 8cm},clip]{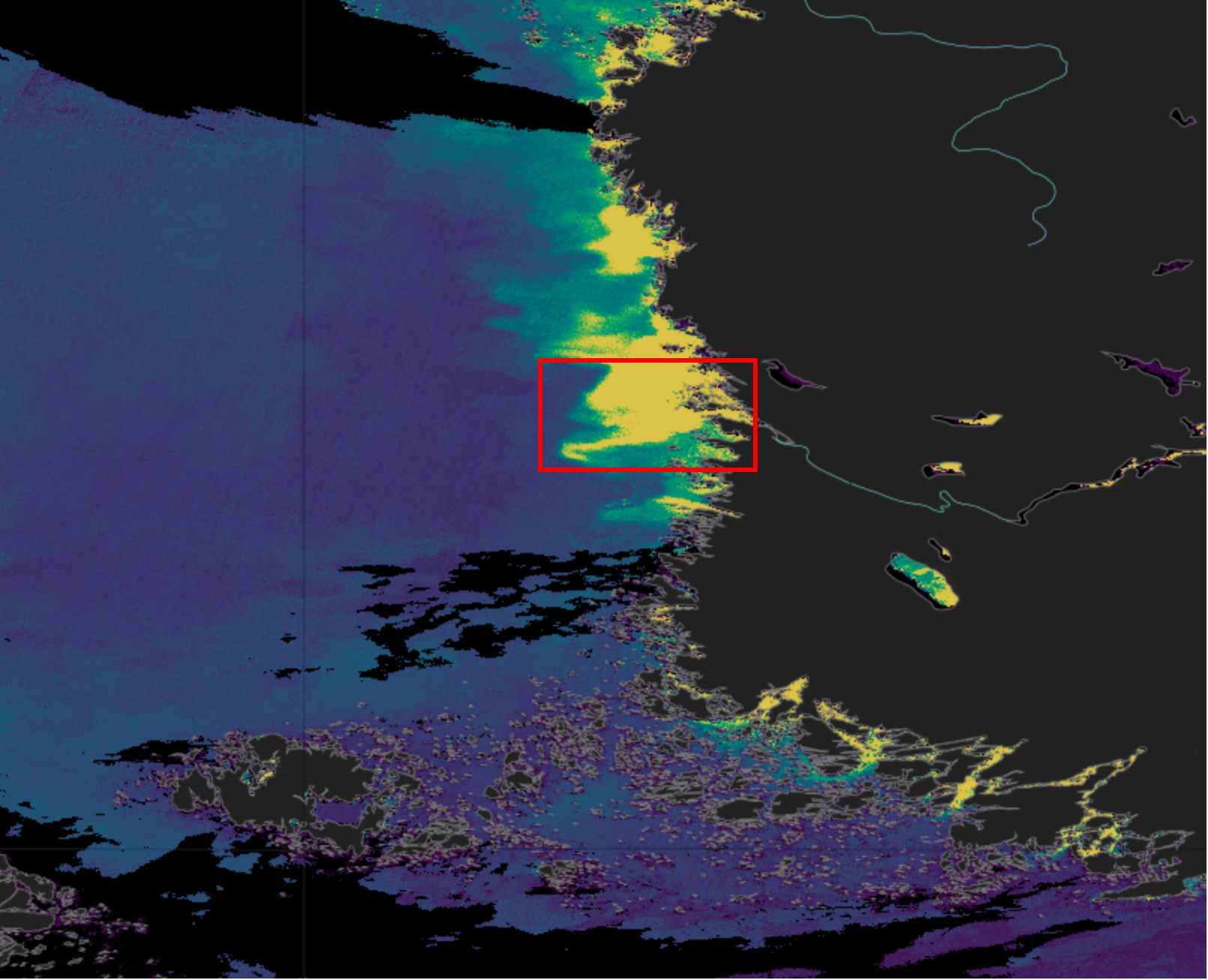} 
    \caption{CMEMS data of {\Chl} concentration in the Baltic Sea (blue-yellow), clouds and cloud coverage (black), and land (dark grey).}
    \label{fig:high_res}
    \vspace{-5mm}
\end{figure}

In this subsection, we introduce the setup for the simulations presented in the following subsections.
For this scenario, we consider the environment illustrated in Fig.~\ref{fig:high_res} in which we will deploy the AUV and track the algal bloom front.
Note that this environment we consider using this data is for simulation purposes only and differs from the satellite data used to inform the GP model, as described in subsection \ref{sub: dataset}.
Here, the {\Chl} concentration is represented by a map that goes gradually from a high concentration in yellow to a low concentration in blue; these are the values that will later be measured by the {\Chl} concentration sensor mounted on the AUV. 
Specifically, the simulated mission will occur inside the red square.
The data used to simulate this environment has a spatial resolution of 300~m by 300~m \cite{CMEMS300m} from the exact satellite data location considered earlier. 
This environment is modeled within the Stonefish simulation environment presented in the previous section. 
Further, the entire {\Chl} concentration map from the satellite image is integrated as a lookup table to enable simulated sampling. 

The source code that implements the algorithm is available as an open-source contribution on two repositories.
The first one is the Gaussian Processes for Adaptive Environmental Sampling (GP4AES) library, which includes the GP model estimator, the gradient estimator, and the motion controller \url{https://github.com/JoanaFonsec/gp4aes}.
The second one is the ROS \cite{ros} interface, which uses the GP4AES library and handles the connection with the AUV's software \url{https://github.com/JoanaFonsec/algalbloom-tracking}.

\begin{figure*}[htpb!]
    \centering
    \includegraphics[width=0.9\textwidth]{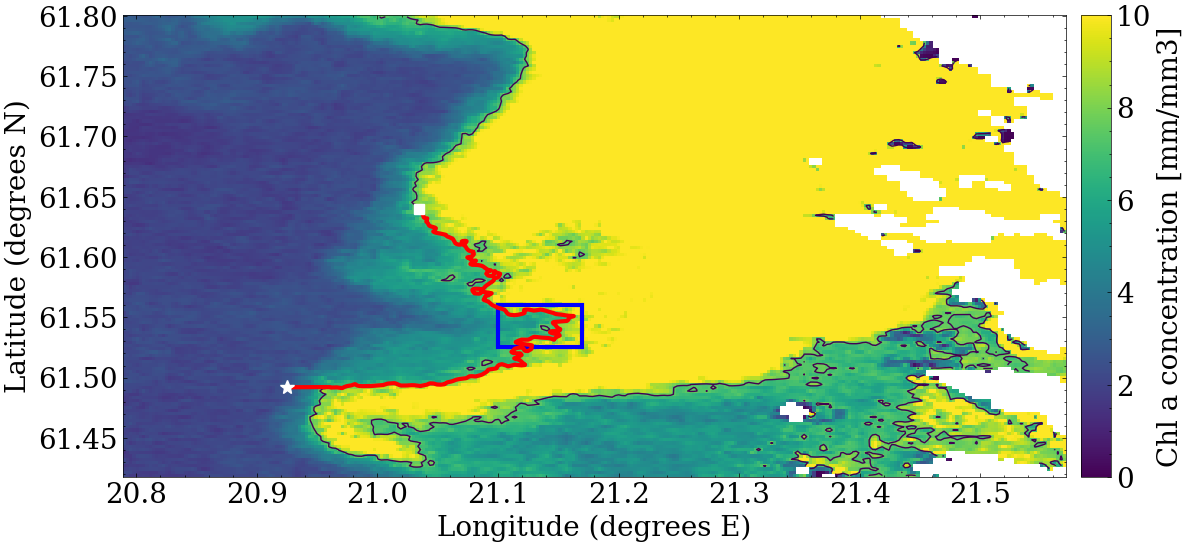}
    \caption{Overview of the full mission having the trajectory of the AUV (red) tracking the front (black) in the {\Chl} field (blue-yellow). The white star indicates the initial position, and the white square the final position.}
    \label{fig:big_map}
    \vspace{-5mm}
\end{figure*}

The simulation starts in the Stonefish Simulator by deploying the AUV close to the front and providing an initial estimated gradient.
When the AUV is near the front, the gradient estimation is triggered. 
The AUV travels at a constant speed of $v=1$~m/s. 
For that reason, we are interested only in the ratio of $\alpha_{\text{seek}}$ and $\alpha_{\text{follow}}$, and thus the latter is set to 1. 
Moreover, based on the available satellite data, we consider $\DeltaRef=7.45$~mg/$\text{m}^3$. 
While tracking the front, the AUV collects measurements at a frequency $f = 1$ Hz while considering a standard deviation of the measurement noise of $\sigma = 10^{-3}$~mg/$\text{m}^3$. 
The measurements are filtered using a weighted moving average filter of size 3, with $w = [0.2, 0.3, 0.5]$
\begin{equation} \label{eq:delta_filtered}
    \delta_{\text{filtered}} (t) = w_{-2}\delta(t-2) + w_{-1}\delta(t-1) + w_{0}\delta(t).
\end{equation}
With the same sampling rate, the gradient is estimated as in \eqref{eq:pred_mean_grad}, using data from the last $n=200$ measurements.  
Then we apply a first-order low pass filter, with $\alpha=0.97$, 
\begin{equation} \label{eq:direction_filtered}
    \nabla\delta_{\text{filtered}}(t) = \alpha\nabla\delta(t-1) + (1-\alpha)\nabla\delta(t).
\end{equation}
The parameters described are summarised in Table~\ref{tab:traj_params}.

\begin{table}[hbtp]
    \centering
    \begin{tabular}{c|c|c|c|c|c}
        $\sigma$ & $\alpha_{\text{seek}}$ & $v$ & $n$ & $\delta_{\text{ref}}$ & $\alpha$ \\ \hline
        $10^{-3}$ mg/$\text{m}^3$ & 10 & 1~m/s & 200 & 7.45~mg/$\text{m}^3$ & 0.97
    \end{tabular}
    \caption{Control algorithm parameters.}
    \label{tab:traj_params}
\end{table}

\subsection{Numerical results}\label{sub: numerical}

In this subsection, we present and analyze the results from simulated missions using two gradient estimation methods.

We illustrate the complete AUV mission in Fig.~\ref{fig:big_map}. The AUV follows the front while collecting {\Chl} concentration measurements, estimating the {\Chl} concentration field and its field gradient, and updating its direction. 
The complete mission has a duration of approximately 23 hours.
The starting position is far from the bloom and represented by the white star, while the final position is on the front and represented by the white square.
In this figure, the AUV closely follows the algal bloom front.
This is further analyzed in subsection \ref{sub: numerical}, in which we focus and zoom-in in on the area inside of the blue square.

\begin{figure*}[htpb!]
    \centering
    \begin{subfigure}{.5\textwidth}
      \centering
      \includegraphics[width=\linewidth]{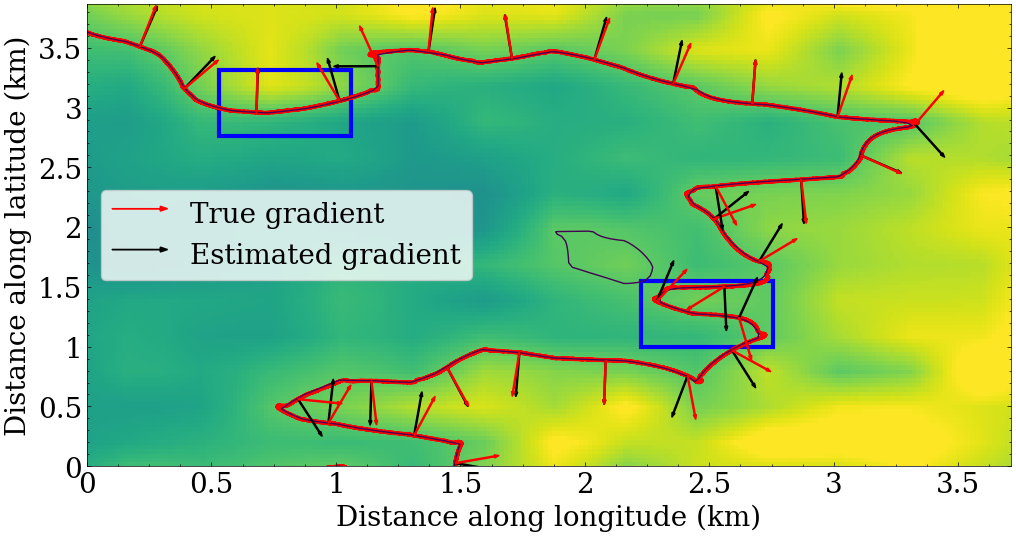}
      \vspace{-5mm}
      \caption{GP estimator}
      \label{fig:GP_zoom1}
    \end{subfigure}%
    \begin{subfigure}{.5\textwidth}
      \centering
      \includegraphics[width=\linewidth]{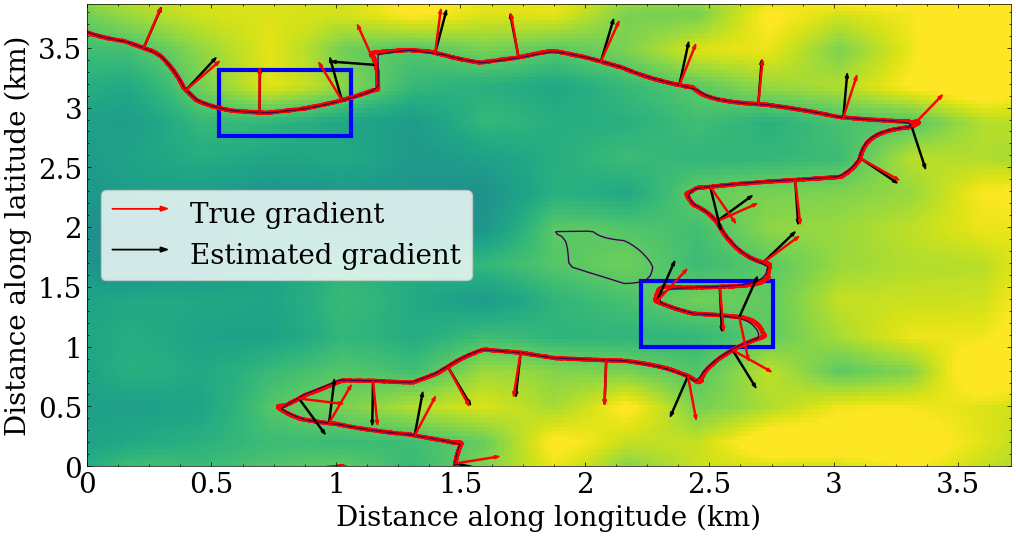}
      \vspace{-5mm}
      \caption{LSQ estimator}
      \label{fig:LSQ_zoom1}
    \end{subfigure}
     \vspace{-2mm}
     \caption{Trajectory of the AUV (red) tracking the front (black line), with arrows representing the true and estimated gradient.}
    \label{fig:zoom1}
    \vspace{-3mm}
\end{figure*}

\begin{figure*}
\centering
\begin{subfigure}{.5\textwidth}
  \centering
  \includegraphics[width=\linewidth]{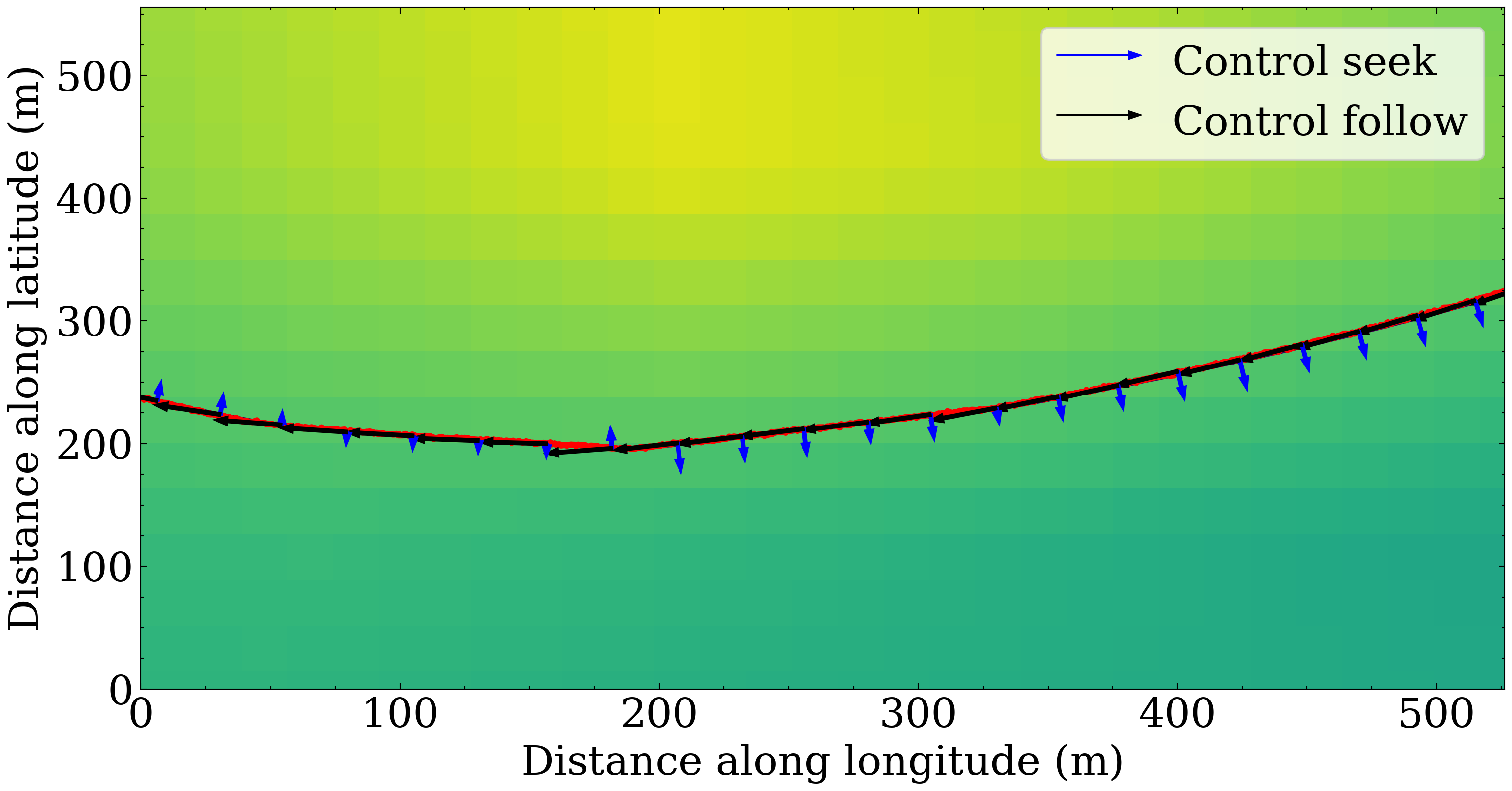}
  \vspace{-5mm}
  \caption{GP estimator}
  \label{fig:GP_zoom2}
\end{subfigure}%
\begin{subfigure}{.5\textwidth}
  \centering
  \includegraphics[width=\linewidth]{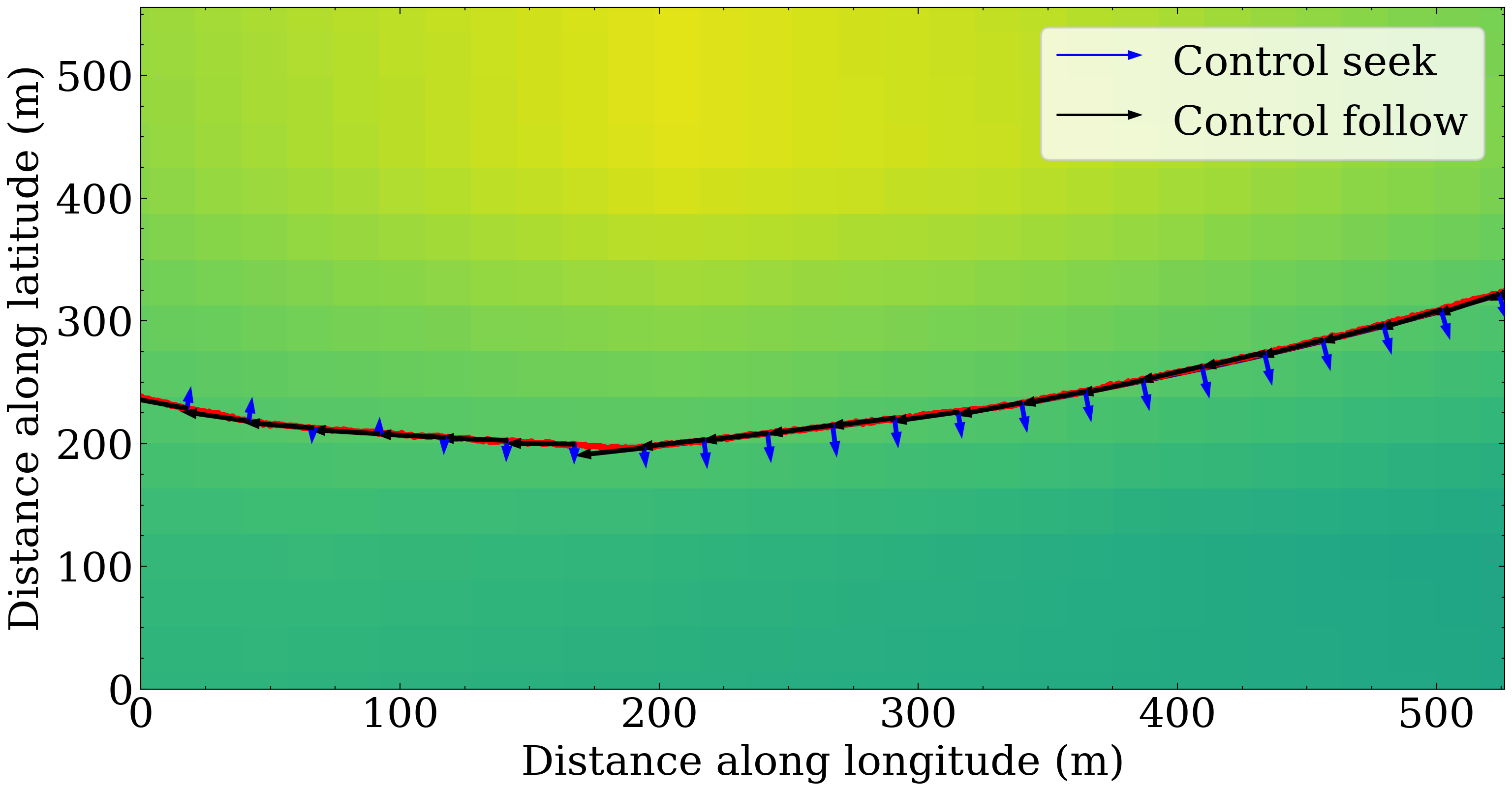}
  \vspace{-5mm}
  \caption{LSQ estimator}
  \label{fig:LSQ_zoom2}
\end{subfigure}
\begin{subfigure}{.5\textwidth}
  \centering
  \includegraphics[width=\linewidth]{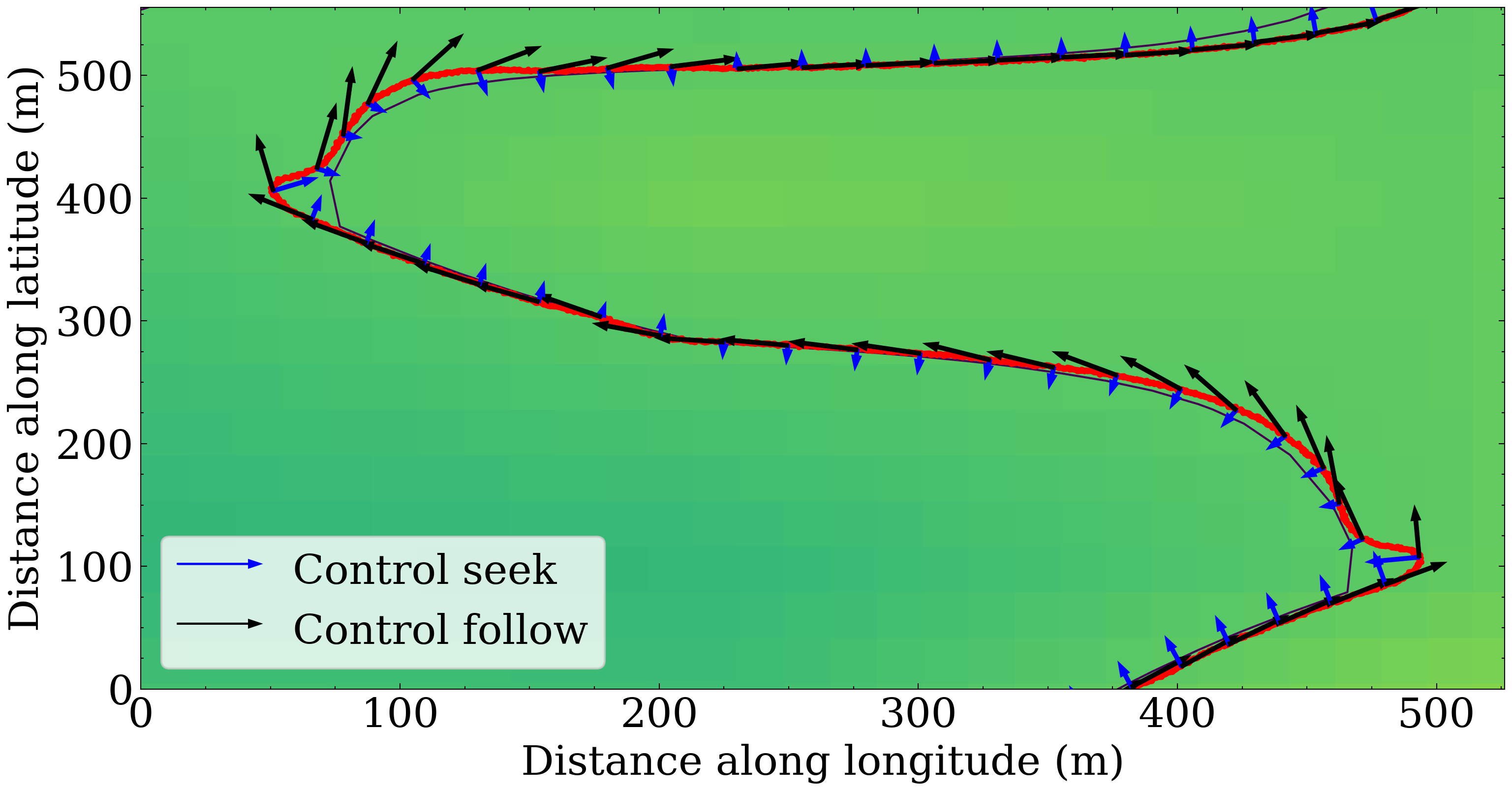}
  \vspace{-5mm}
  \caption{GP estimator}
  \label{fig:GP_zoom3}
\end{subfigure}%
\begin{subfigure}{.5\textwidth}
  \centering
  \includegraphics[width=\linewidth]{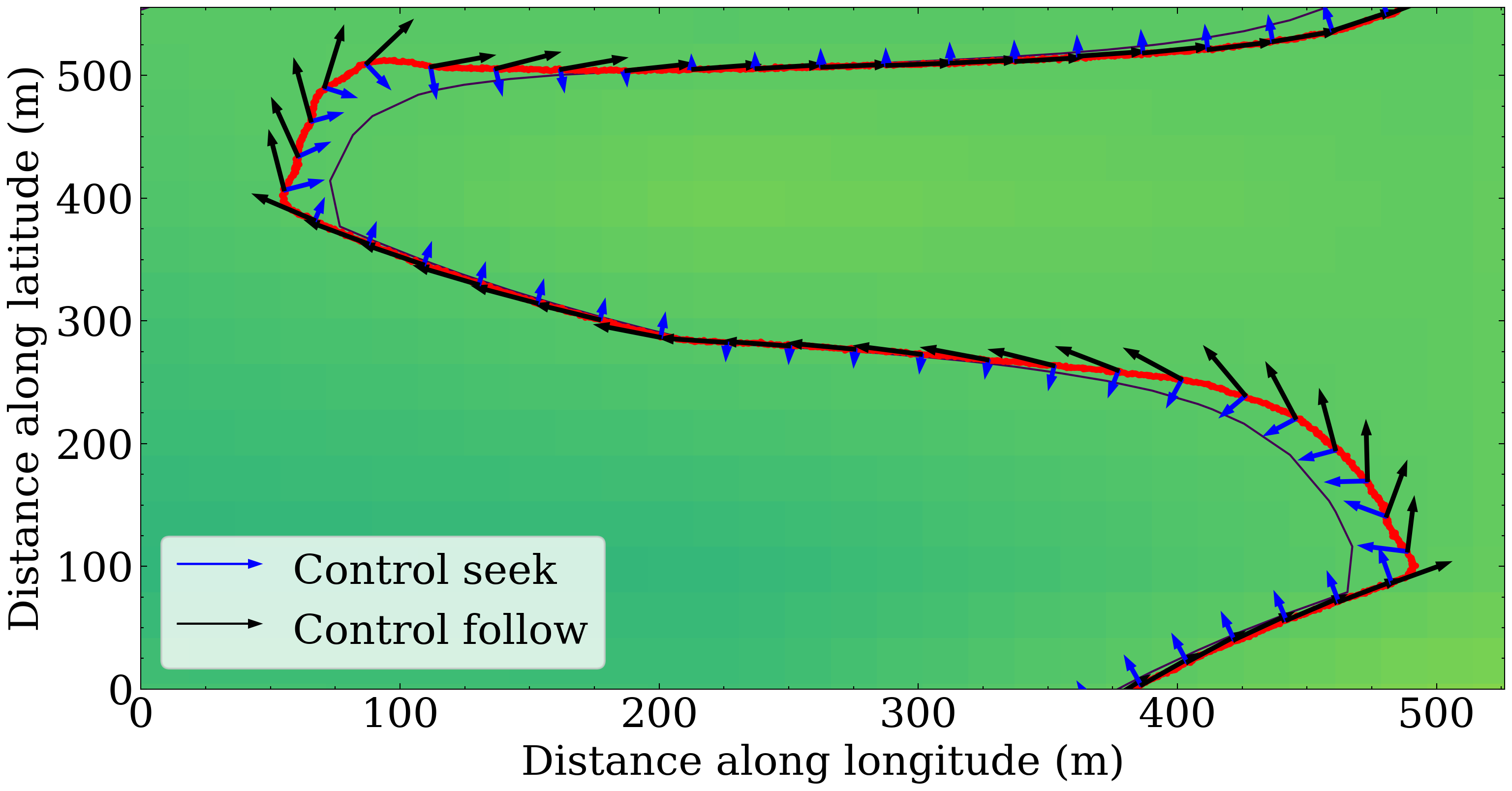}
  \vspace{-5mm}
  \caption{LSQ estimator}
  \label{fig:LSQ_zoom3}
\end{subfigure}
\vspace{-2mm}
    \caption{AUV path (red) tracking the front (black), with arrows representing seek and follow components of the control law.}
\label{fig:zoom2e3}
\vspace{-6mm}
\end{figure*}

\begin{figure*}[btp]
    \centering
    \begin{subfigure}{.5\textwidth}
      \centering
      \includegraphics[width=\linewidth]{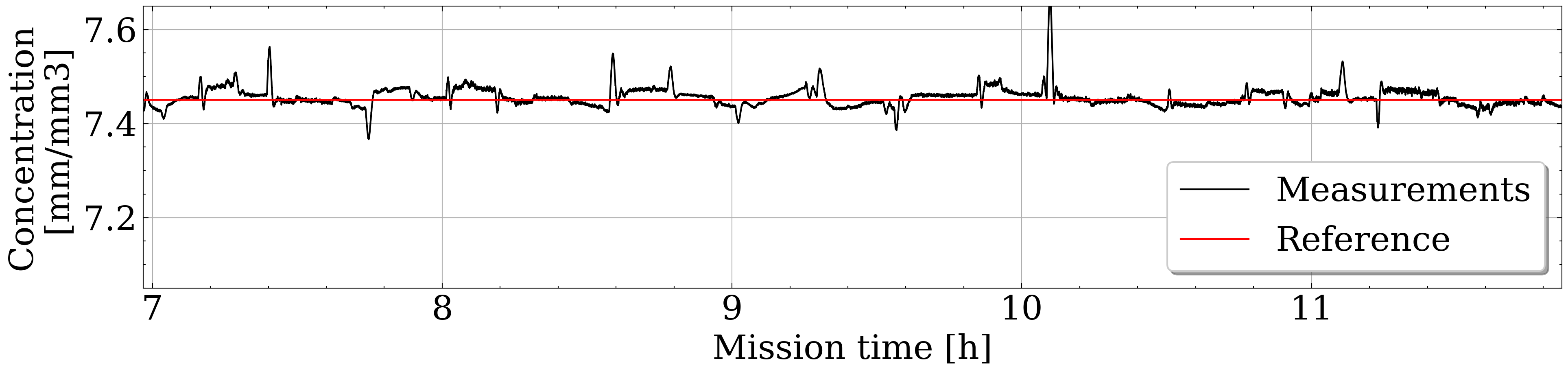}
      \vspace{-5mm}
      \caption{GP estimator}
      \label{fig:GP_chl}
    \end{subfigure}%
    \begin{subfigure}{.5\textwidth}
      \centering
      \includegraphics[width=\linewidth]{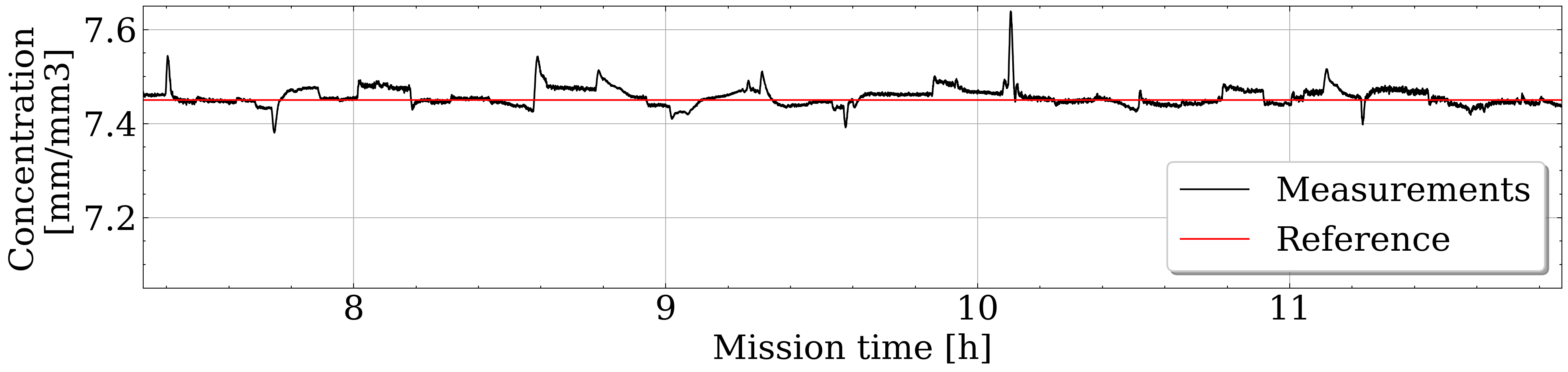}
      \vspace{-5mm}
      \caption{LSQ estimator}
      \label{fig:LSQ_chl}
    \end{subfigure}
     \vspace{-2mm}
     \caption{Concentration of {\Chl}: measurements from the AUV, and reference value.}
    \label{fig:chl}
    \vspace{-2mm}
\end{figure*}

\begin{figure*}[htpb!]
    \centering
    \begin{subfigure}{.5\textwidth}
      \centering
      \includegraphics[width=\linewidth]{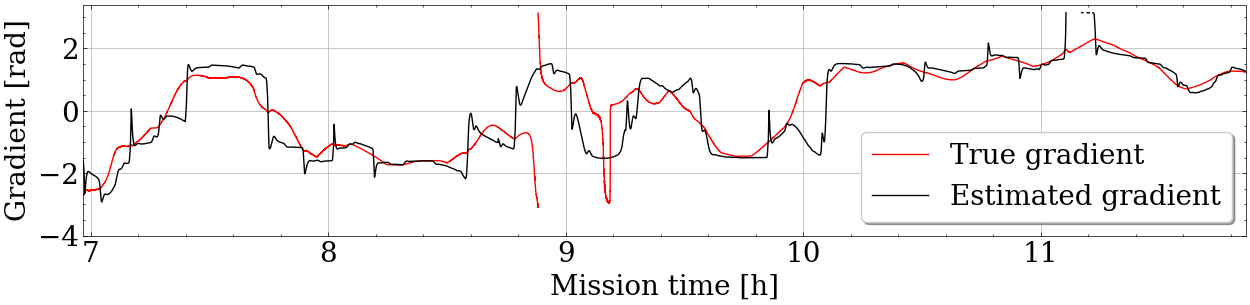}
      \vspace{-5mm}
      \caption{GP estimator}
      \label{fig:GP_gradient}
    \end{subfigure}%
    \begin{subfigure}{.5\textwidth}
      \centering
      \includegraphics[width=\linewidth]{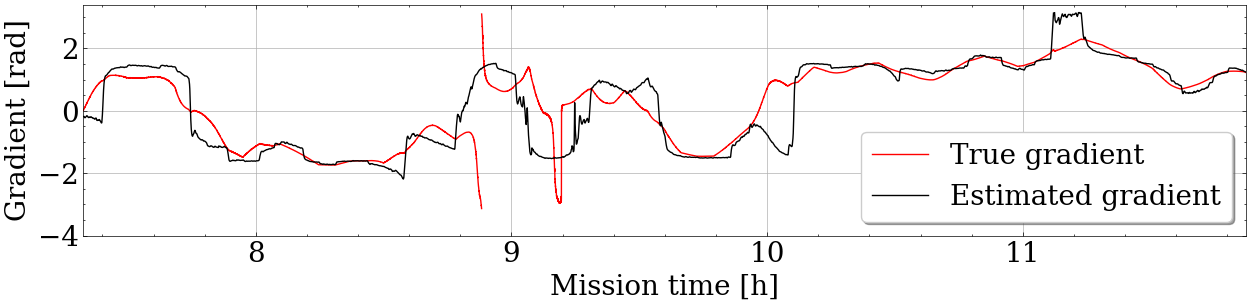}
      \vspace{-5mm}
      \caption{LSQ estimator}
      \label{fig:LSQ_gradient}
    \end{subfigure}
     \vspace{-2mm}
     \caption{Gradient of {\Chl}: AUV estimated gradient, and true gradient.}
    \label{fig:gradient}
    \vspace{-2mm}
\end{figure*}

\begin{figure*}[htpb!]
    \centering
    \begin{subfigure}{.5\textwidth}
      \centering
      \includegraphics[width=\linewidth]{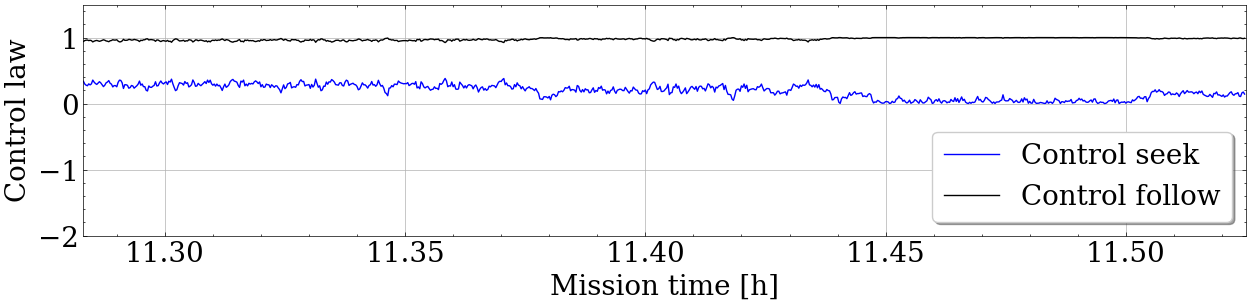}
      \vspace{-5mm}
      \caption{GP estimator}
      \label{fig:GP_control1}
    \end{subfigure}%
    \begin{subfigure}{.5\textwidth}
      \centering
      \includegraphics[width=\linewidth]{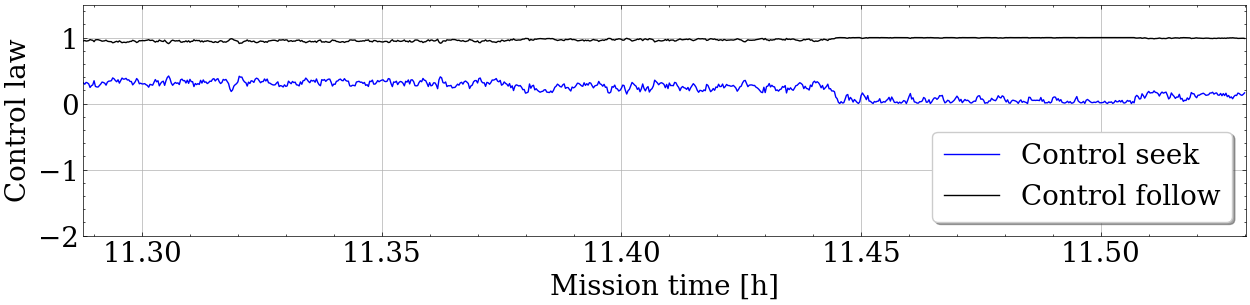}
      \vspace{-5mm}
      \caption{LSQ estimator}
      \label{fig:LSQ_control1}
    \end{subfigure}
        \begin{subfigure}{.5\textwidth}
      \centering
      \includegraphics[width=\linewidth]{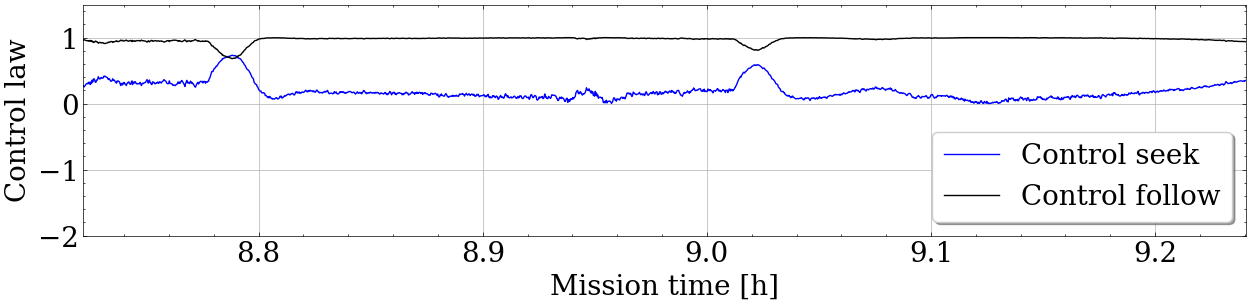}
      \vspace{-5mm}
      \caption{GP estimator}
      \label{fig:GP_control2}
    \end{subfigure}%
    \begin{subfigure}{.5\textwidth}
      \centering
      \includegraphics[width=\linewidth]{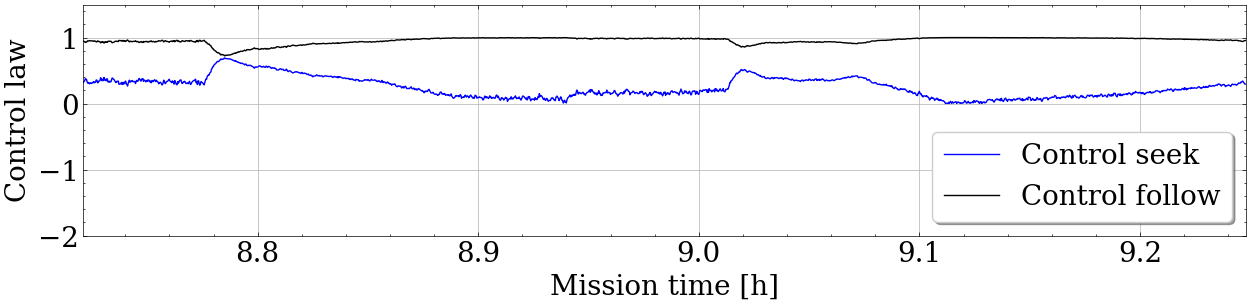}
      \vspace{-5mm}
      \caption{LSQ estimator}
      \label{fig:LSQ_control2}
    \end{subfigure}
     \vspace{-2mm}
     \caption{Control law components: seek and follow.}
    \label{fig:controls}
    \vspace{-5mm}
\end{figure*}

In Fig.~\ref{fig:zoom1}, we zoom in on a region of the longer mission, previously defined by a blue square, to focus on the performance of the gradient estimation and front tracking algorithm.
This region corresponds to about 5 hours of mission time.
Here we illustrate the gradient performance through arrows representing the true and estimated gradients along the path.
The true gradient refers to the gradient that the AUV would be able to compute if it had access to the global information of the field. 
We compute it by taking the spatial derivative of the {\Chl} concentration field.
The estimated gradient refers to the output of the gradient estimator, as in \eqref{eq:gradient}.
The angle between the true and estimated gradient arrows indicates the gradient error. 
However, in this scenario, the {\Chl} concentration field is non-convex and fast-changing even in small areas. 
Therefore, the gradient is an abstraction that gives an idea of direction rather than an exact measure of the gradient. 
If we analyze the straight portions of the path, we would say that the error is very close to zero. In contrast, we could say that the error is larger by analyzing the portions of the path with higher curvature. At the same time, the gradient looks ambiguous and sensitive to small changes in position.
As for both the control performance and comparison between gradient estimators, this figure doesn't allow for such analysis so we zoom-in in on the two areas inside the blue squares.

Let us now analyze the performance of the control and its control components in Fig.~\ref{fig:zoom2e3}.
These figures correspond to the two zoom-in locations in the previous figure; the front is the thin black line, and a thicker red line represents the AUV path.
We also plot the \textit{seek} and \textit{follow} components of the control law using arrows along the AUV path.
The control law is constructed as in eq.~\ref{eqn:law}, and it is a sum of the seek component, which has the same direction as the estimated gradient, and the follow component, which has a perpendicular direction with respect to the estimated gradient.
This sum constitutes the control law corresponding to the AUV's direction of movement.
The first zoom-in corresponds to about 15 minutes of mission time.
Here, for both estimators in Fig.~\ref{fig:GP_zoom2} and Fig.~\ref{fig:LSQ_zoom2}, the AUV always follows the front closely and with minimal error and without visible differences of performance among the estimators.
This is expected as the front is smooth in curvature, and the AUV remains on top of the front. The control seek component accounts for small adjustments in the trajectory.
The second zoom-in corresponds to about 30 minutes of mission time.
In Fig.~\ref{fig:GP_zoom3} and Fig.~\ref{fig:LSQ_zoom3}, the AUV remains on top of the front most of the time; thus, the control follow component dominates the control law. 
On the other hand, once the curvature changes faster, the control follow component is reduced, and the control seek component becomes the dominating component.
In this scenario of fast-changing curvature, the AUV seems to have a delay in updating its direction.
Two leading causes for this behavior are the AUV's turning radius and the update function with the update rate on the gradient. 
The gradient's update function in \eqref{eq:direction_filtered} introduces a delay and a cut-off frequency. 
This cut-off frequency is inversely proportional to the update rate. Hence, the algorithm's performance becomes a trade-off between the smoothness introduced by the update function with a lower update rate and the delay it introduces.
For this scenario, we considered smoothness of movement a more important objective than the apparent delay in the tighter curvature of the front. 
As for the performance comparison between the two gradient estimation methods, it is apparent from Fig.~\ref{fig:zoom2e3} that the GP gradient estimator allows for closer tracking of the front, most notably at regions with fast-changing curvature.

We further analyze the algorithm's behavior through time series plots in Fig.~\ref{fig:chl} and Fig.~\ref{fig:gradient} corresponding to the zoom-in area in Fig.~\ref{fig:zoom1}.
First, we consider the {\Chl} concentration measurements taken along the path in Fig.~\ref{fig:chl}.
The time series indicates that, as seen in the previous figures, the AUV is always on top of the front, oscillating around it and with a minimal error, in this case, lower than $\pm 0.1\text{mg/m}^3$. 
Second, we consider the gradient field estimation also taken along the mission, in Fig.~\ref{fig:gradient}.
This time series also confirms what we saw in the previous figures. Here we can see both the delay of the estimated gradient and its smoothness compared with the true gradient.
No major performance differences exist among the gradient estimators for either of the time series plots.
Finally, let us further analyze the control law, considering the time series of the two control components in Fig.~\ref{fig:controls}, for the regions defined in Fig.~\ref{fig:zoom2e3}.
In Fig.~\ref{fig:GP_control1} and Fig.~\ref{fig:LSQ_control1}, we consider the first zoom-in area with an almost linear segment of the front.
Here we get a near-constant behavior of both controller components, where the follow component is almost always 1, and the seek component is near zero most of the time.
In Fig.~\ref{fig:GP_control2} and Fig.~\ref{fig:LSQ_control2}, we consider the second zoom-in, which contains two tight curves.
Here the follow component also dominates the control law, with exceptions in two instances at $t=8.79$ and $t=9.02$, corresponding to the two peaks in trajectory curvature, in Fig.~\ref{fig:GP_zoom3} and Fig.~\ref{fig:LSQ_zoom3}.
For both examples, the follow component dominates the control law, and the seek component increases when the AUV is far from the front.
Here, the difference in performance of the two gradient estimators is not visible.
Thus, the following subsection is dedicated to properly evaluating the behavior of the different estimators.

\subsection{Sensitivity analysis}\label{sub: comparison}

In this subsection, we analyze the performance of the estimation algorithms for a varying standard deviation of {\Chl} concentration  sensor noise.

The sensor noise we consider for this analysis varies between $10^{-3}\text{mg/m}^3$ and $10^{-1}\text{mg/m}^3$. 
This is in line with the {\Chl} concentration sensors in the market, which have a resolution of $10^{-2}\text{mg/m}^3$.
To analyze the impact of the noise introduced by the sensor, we remove the moving average window used on the measurements introduced by eq.~\ref{eq:delta_filtered}.
We then run multiple simulations using different standard deviations of the sensor noise and estimation methods.
For this analysis, we consider two estimation methods: Gaussian Process regression, which is used in this paper, and Least Squares, which is used in the literature.
We obtain both the gradient estimation and tracking errors from each simulation. 
Each simulation results in a data point in this subsection's results.

\begin{figure} [btp] 
    \centering    \includegraphics[width=.49\textwidth]{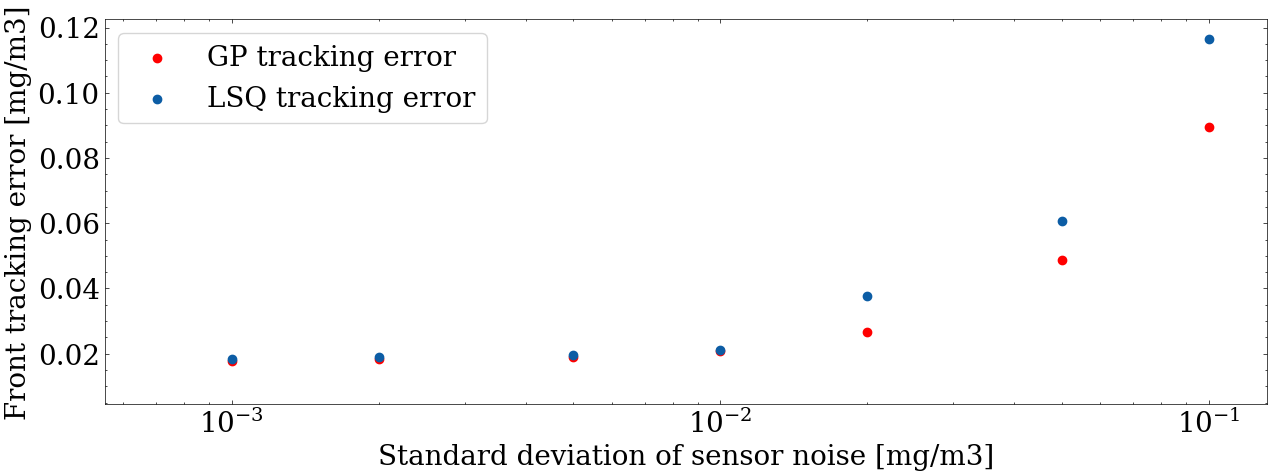} 
    \caption{Influence of sensor noise in the tracking error, for two different estimation algorithms: GP and LSQ.}
    \label{fig:chl_error}
    \vspace{-3mm}
\end{figure}

The gradient estimation error corresponds to the difference between the true gradient $\nabla\delta(t)$ and the estimated gradient $\nabla\bar{\delta}(t)$.
In Fig.~\ref{fig:chl_error}, we illustrate the impact of sensor noise on this error.
Here the tracking error increases exponentially with the sensor noise for both estimators.
The difference in performance between estimators becomes exponentially bigger from the standard deviation of sensor noise of $0.02\text{mg/m}^3$. 

\begin{figure} [btp] 
    \centering    \includegraphics[width=.49\textwidth]{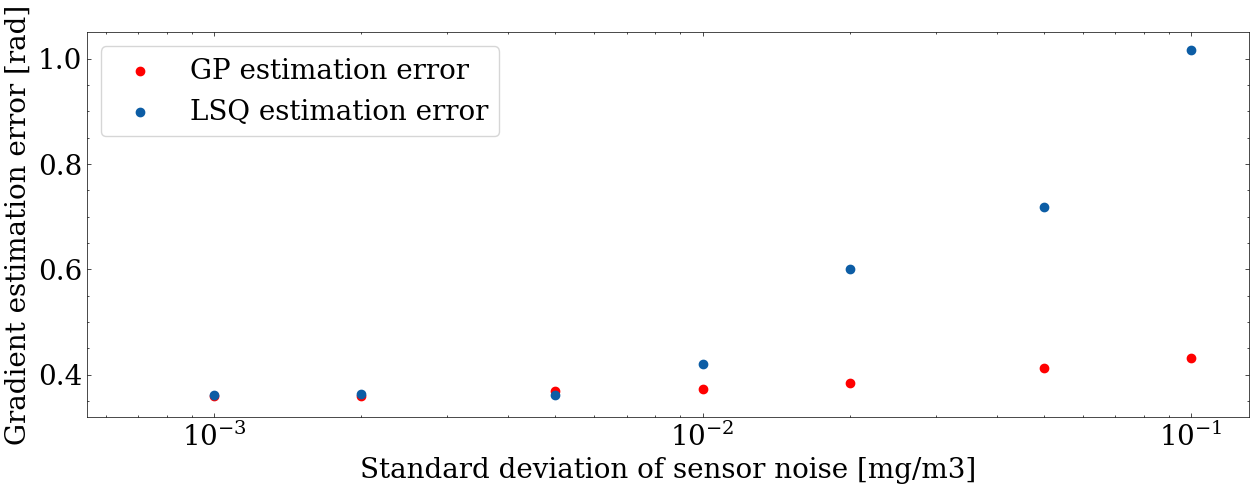} 
    \caption{Influence of sensor noise in the gradient estimation error, for two different estimation algorithms: GP and LSQ.}
    \label{fig:gradient_error}
    \vspace{-5mm}
\end{figure}

The tracking error corresponds to the difference between the measured {\Chl} concentration $\delta(t)$ and the {\Chl} reference $\DeltaRef$.
In Fig.~\ref{fig:gradient_error}, we illustrate the impact of sensor noise on this error.
Here the tracking error increases exponentially with the sensor noise for both estimators (note that the x-axis is in logarithmic scale).
The difference in performance between estimators becomes much bigger from the standard deviation of sensor noise of $0.01\text{mg/m}^3$. 

From this comparison, we can conclude that the GP estimator performs better than the LSQ-based estimator for gradient estimation and front tracking. 
We can also conclude that the error grows substantially from a standard deviation of sensor noise of $0.01\text{mg/m}^3$.

\section{Experimental results in the Baltic Sea}\label{s: experiments}

\begin{figure} [btp] 
    \centering    
    \includegraphics[width=.49\textwidth,trim={0 21cm 0 2cm},clip]{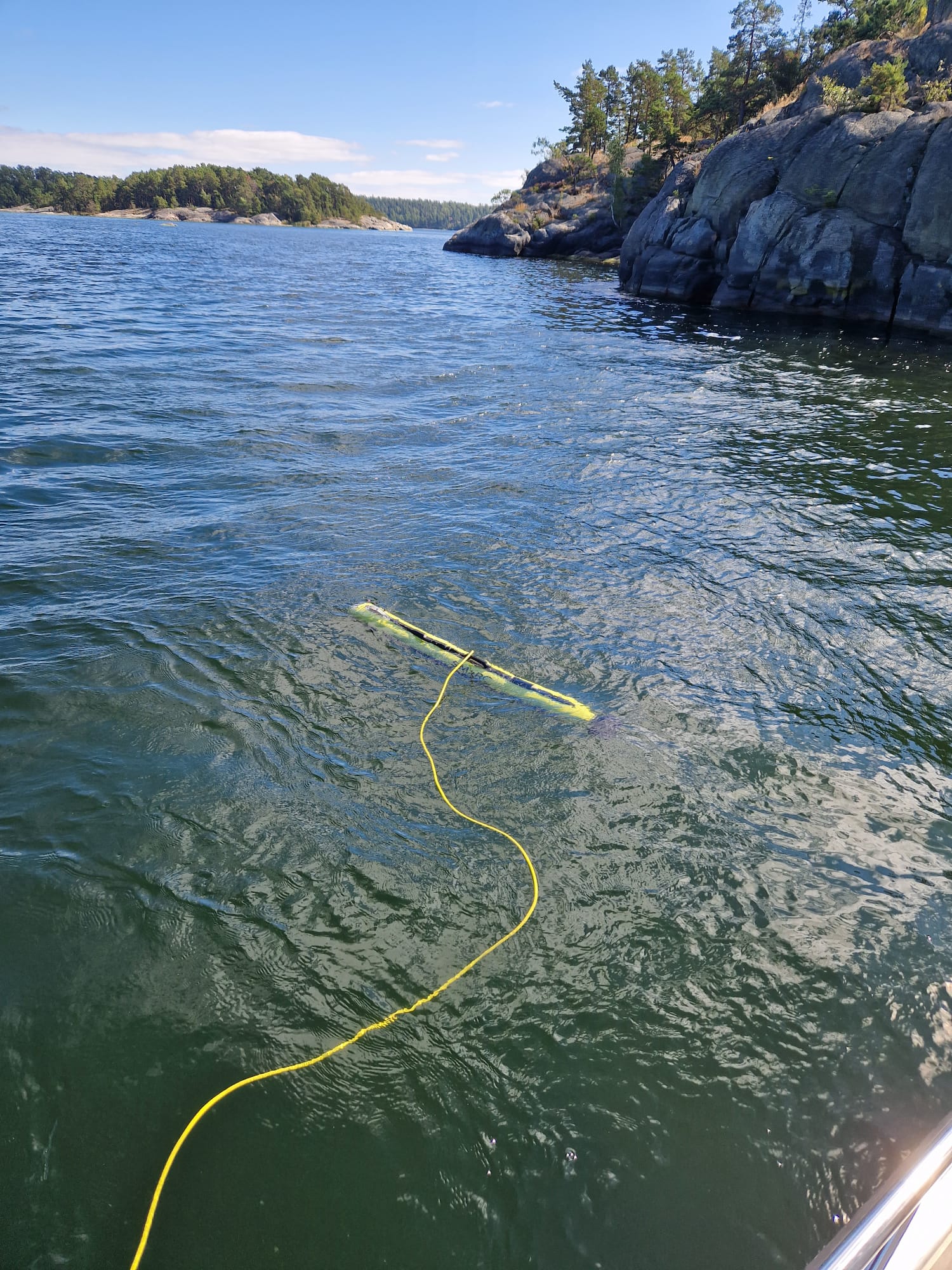} 
    \caption{The AUV in the water on the mission day, while tethered to the boat.}
    \label{fig:water_sam}
    \vspace{-5mm}
\end{figure}

This section contains the results obtained from experiments using the components introduced in the previous sections, summarised in Fig.~\ref{fig:sam_sys_diag}.
Along with the numerical simulation study, the gradient estimator and motion controller were validated in field experiments. 
These experiments took place in the Stockholm archipelago in the Baltic Sea, near the island of Djur{\"o}, in an area, with slightly different starting positions.
We ensured safety by following the AUV by boat as in Fig.\ref{fig:water_sam}, by reducing the AUV's speed to approximately $0.1\text{m/s}$, and by constraining the survey area to a small region with low ferry traffic. 
We also ran the surveys on the surface so that it is possible to get position feedback using GPS, thereby counteracting the uncertainty from exclusively using dead reckoning. 
To satisfy the constraints on maximum speed and small survey area, we scaled the data in Fig.\ref{fig:high_res} 100 times.
This data was used to simulate the {\Chl} concentration sensor as a look-up table, given that there are no algal blooms in this small area.

\begin{figure} [btp]
  \begin{subfigure}[b]{\linewidth}
        \includegraphics[width=\linewidth]{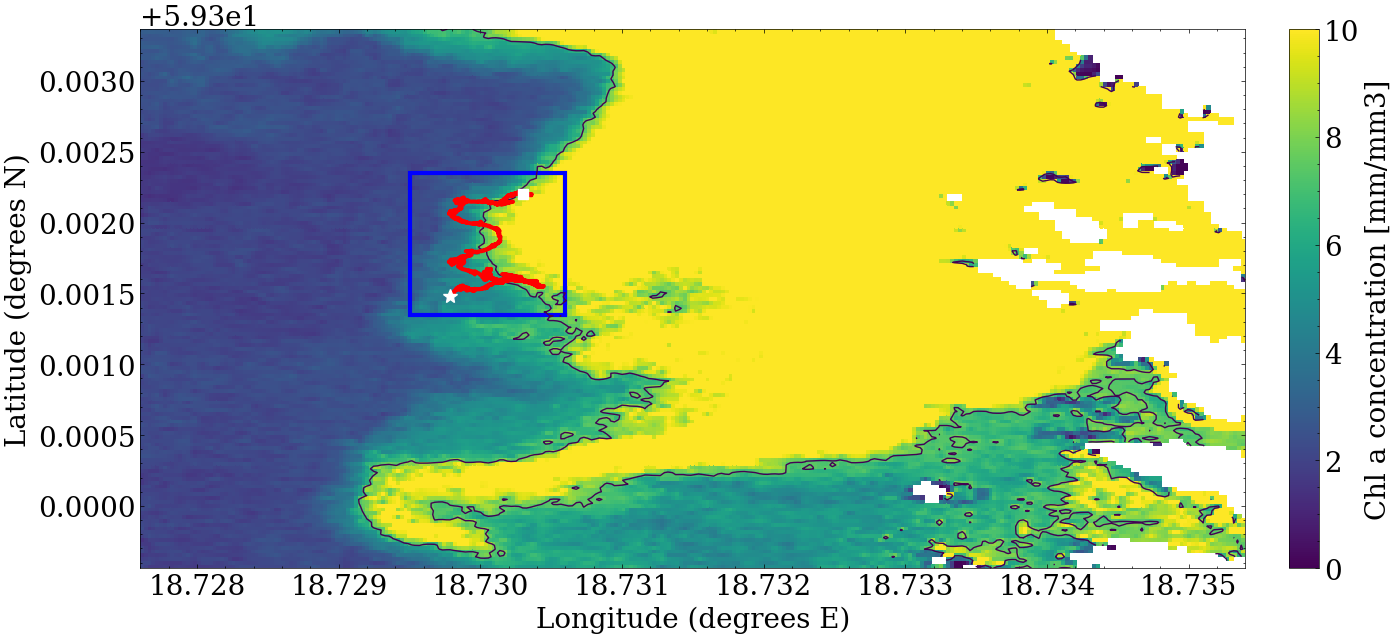}
        \vspace{-5mm}
        \caption{Survey in the 18th of July 2022.}
        \label{fig:experiment_July_bigmap}
  \end{subfigure}\\ 
  \begin{subfigure}[b]{\linewidth}
        \includegraphics[width=\linewidth]{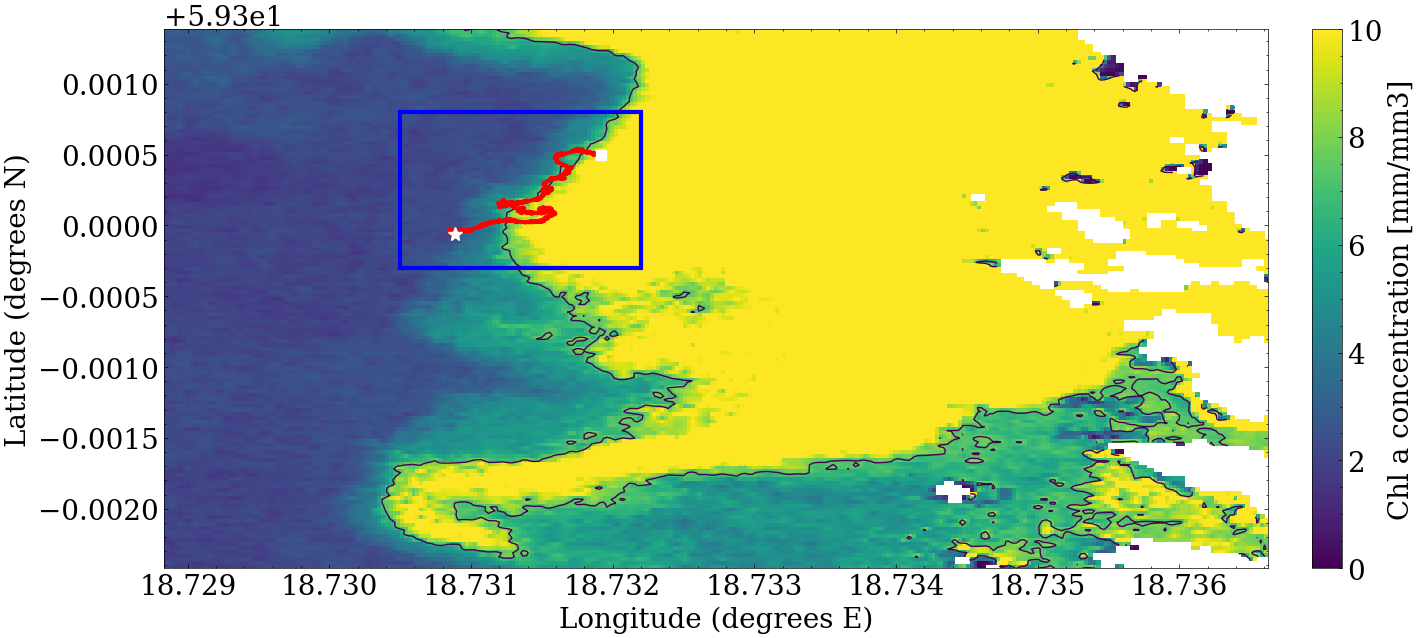}
        \vspace{-5mm}
        \caption{Survey in the 11th of August 2022.}
        \label{fig:experiment_Aug_bigmap}
    \end{subfigure}\\
      \begin{subfigure}[b]{\linewidth}
        \includegraphics[width=\linewidth]{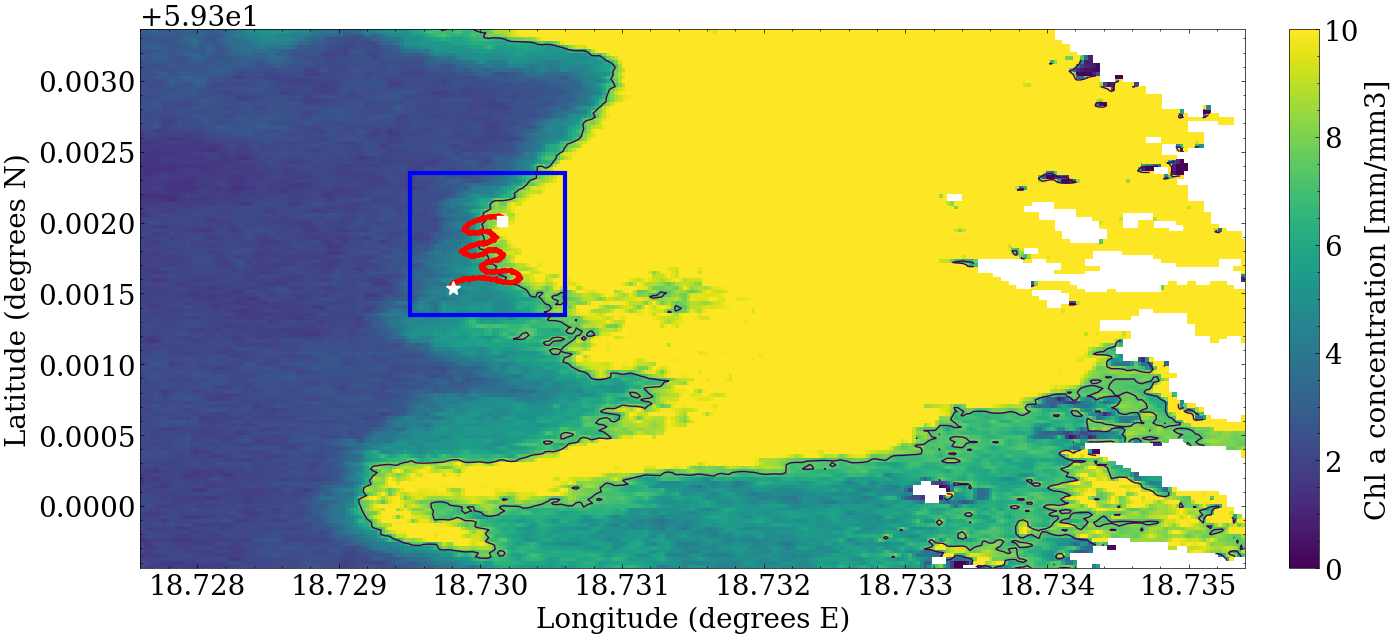}
        \vspace{-5mm}
        \caption{Simulation under experiment conditions.}
        \label{fig:scaled_bigmap}
    \end{subfigure}\\
  \vspace{-5mm}
    \caption{Two experimental surveys and a simulated scenario under experimental conditions with AUV trajectory (red) tracking the front (black line) in the {\Chl} map (blue-yellow).}
    \label{fig:experiments_overview}
    \vspace{-5mm}
\end{figure}

In Fig.~\ref{fig:experiments_overview}, we illustrate two surveys and a simulated scenario. 
The first survey, in Fig.~\ref{fig:experiment_July_bigmap}, was conducted on July 18th, 2022. 
It corresponds to about 12 minutes of mission time at an average speed of $0.11\text{m/s}$.
The second survey, in Fig.~\ref{fig:experiment_Aug_bigmap}, was conducted on August 11th, 2022.
It corresponds to about 10 minutes of mission time at an average speed of $0.11\text{m/s}$.
To better analyze and understand the results, we also present a simulated scenario Fig.~\ref{fig:scaled_bigmap} in which we set all the conditions to match the conditions of the experiments as closely as possible.
These include the scaled data as a look-up table for the simulated sensor, the initial position, GPS noise, and a similar survey length of approximately 10 minutes.

\begin{figure*}[htpb!]
    \centering
    \begin{subfigure}{.3\textwidth}
      \centering
      \includegraphics[width=\linewidth]{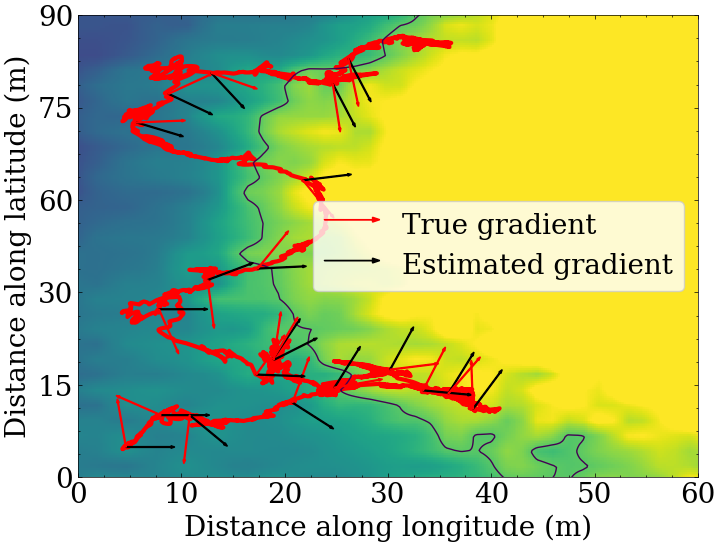}
      \vspace{-5mm}
      \caption{Survey in the 18th of July 2022.}
      \label{fig:experiment_July_zoom1}
    \end{subfigure}%
    \begin{subfigure}{.38\textwidth}
      \centering
      \includegraphics[width=\linewidth]{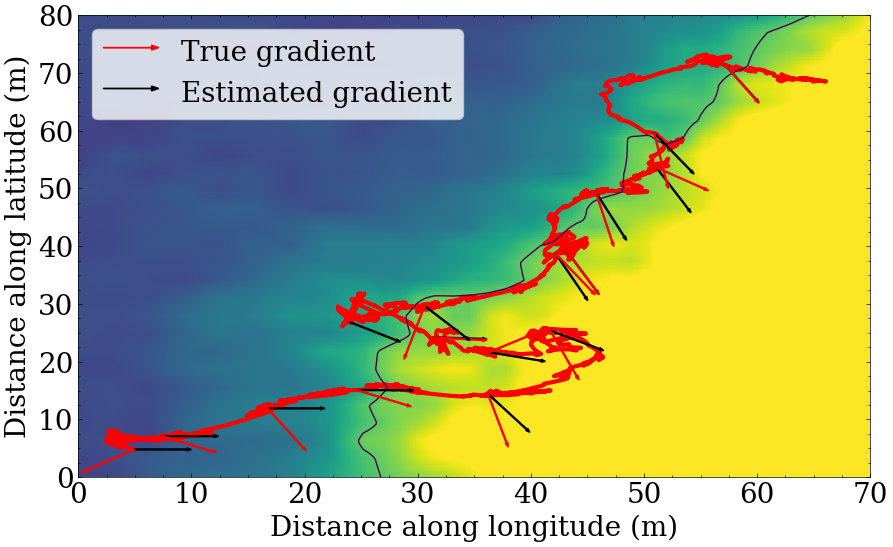}
      \vspace{-5mm}
      \caption{Survey in the 11th of August 2022.}
      \label{fig:experiment_Aug_zoom1}
    \end{subfigure}
    \begin{subfigure}{.3\textwidth}
      \centering
      \includegraphics[width=\linewidth]{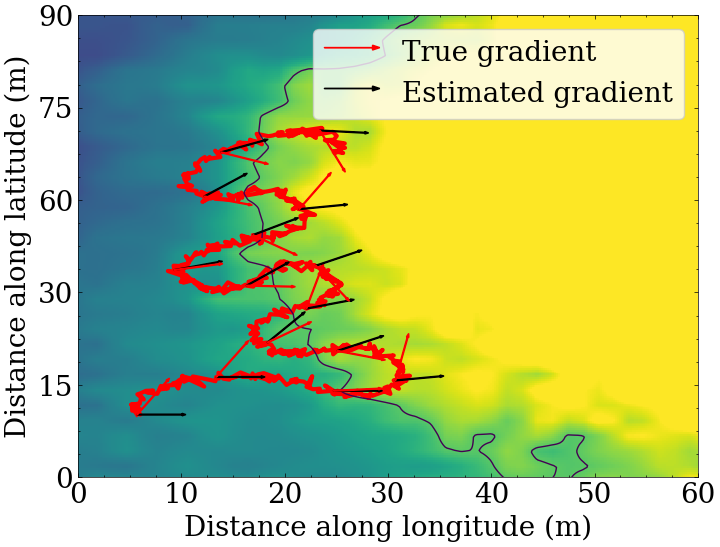}
      \vspace{-5mm}
      \caption{Simulation under experiment conditions.}
      \label{fig:scaled_zoom1}
    \end{subfigure}
     \vspace{-2mm}
     \caption{Trajectory of the AUV (red) tracking the front (black line), with arrows representing the true and estimated gradient.}
    \label{fig:exp_zoom1}
    \vspace{-3mm}
\end{figure*}

Let us now analyze the results by zooming in on the survey areas and examining the components of our algorithm.
In the first zoom-in in Fig.~\ref{fig:exp_zoom1}, we evaluate the trajectory of the AUV and gradient estimation.
The trajectory of the real AUV in both surveys, in Fig.~\ref{fig:experiment_July_zoom1} and Fig.~\ref{fig:experiment_Aug_zoom1}, appears very jittery as opposed to previous chapters in which we simulated a mission. 
This effect is due to two phenomena: the noise in the GPS signal and water currents. 
Water currents drag the AUV to move in a different direction than the one the algorithm calculated.
These currents are also one of the causes for the jittery GPS signal as, in the case of our AUV, the GPS receiver is a few centimeters above the surface, and the signal quality is directly influenced by the existence of waves that can partially cover the signal.
In Fig.~\ref{fig:scaled_zoom1}, we attempted to emulate this phenomenon by introducing Gaussian noise on the GPS receiver.

\begin{figure*}[htpb!]
    \centering
    \begin{subfigure}{.33\textwidth}
      \centering
      \includegraphics[width=\linewidth]{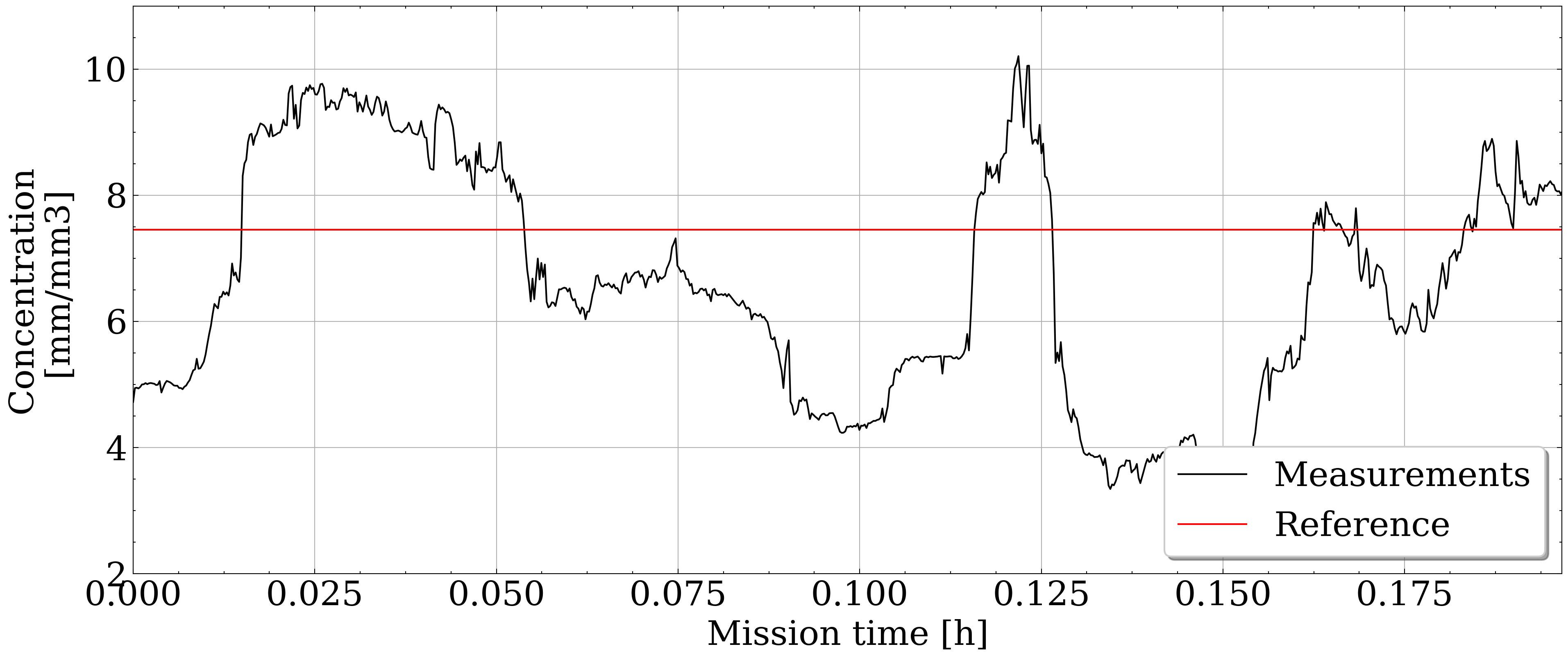}
      \vspace{-5mm}
      \caption{Survey in the 18th of July 2022.}
      \label{fig:experiment_July_chl}
    \end{subfigure}%
    \begin{subfigure}{.33\textwidth}
      \centering
      \includegraphics[width=\linewidth]{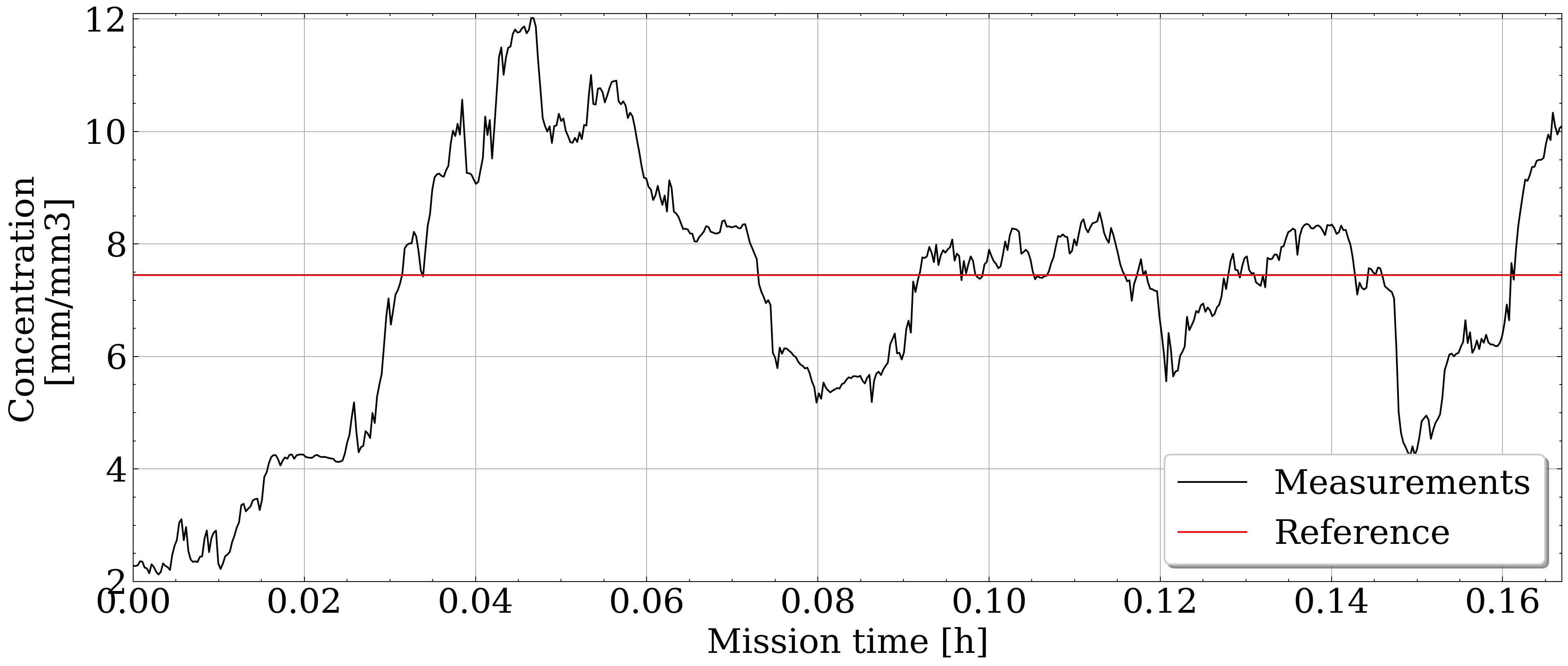}
      \vspace{-5mm}
      \caption{Survey in the 11th of August 2022.}
      \label{fig:experiment_Aug_chl}
    \end{subfigure}
    \begin{subfigure}{.33\textwidth}
      \centering
      \includegraphics[width=\linewidth]{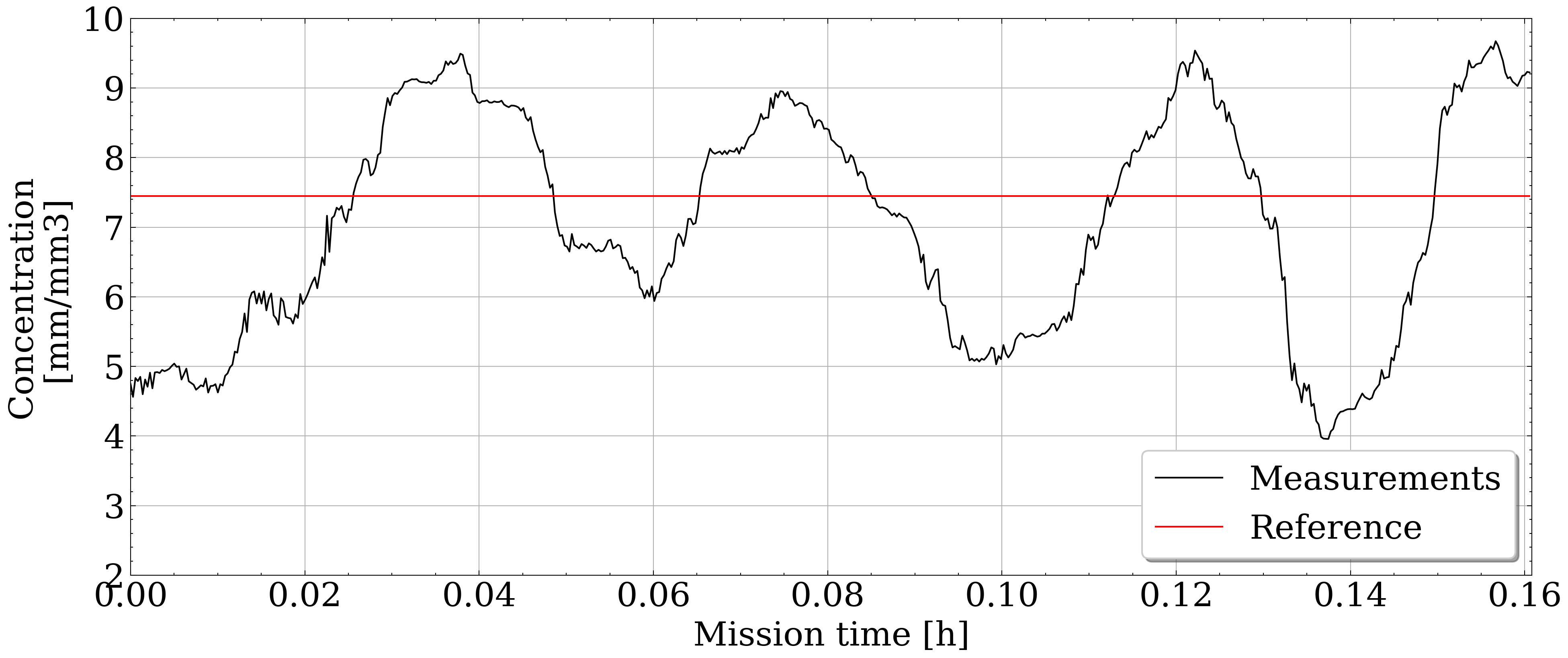}
      \vspace{-5mm}
      \caption{Simulation under experiment conditions.}
      \label{fig:scaled_chl}
    \end{subfigure}
     \vspace{-2mm}
     \caption{Concentration of chlorophyll a: measurements from the AUV, and reference value.}
    \label{fig:exp_chl}
    \vspace{-3mm}
\end{figure*}

\begin{figure*}[htpb!]
    \centering
    \begin{subfigure}{.33\textwidth}
      \centering
      \includegraphics[width=\linewidth]{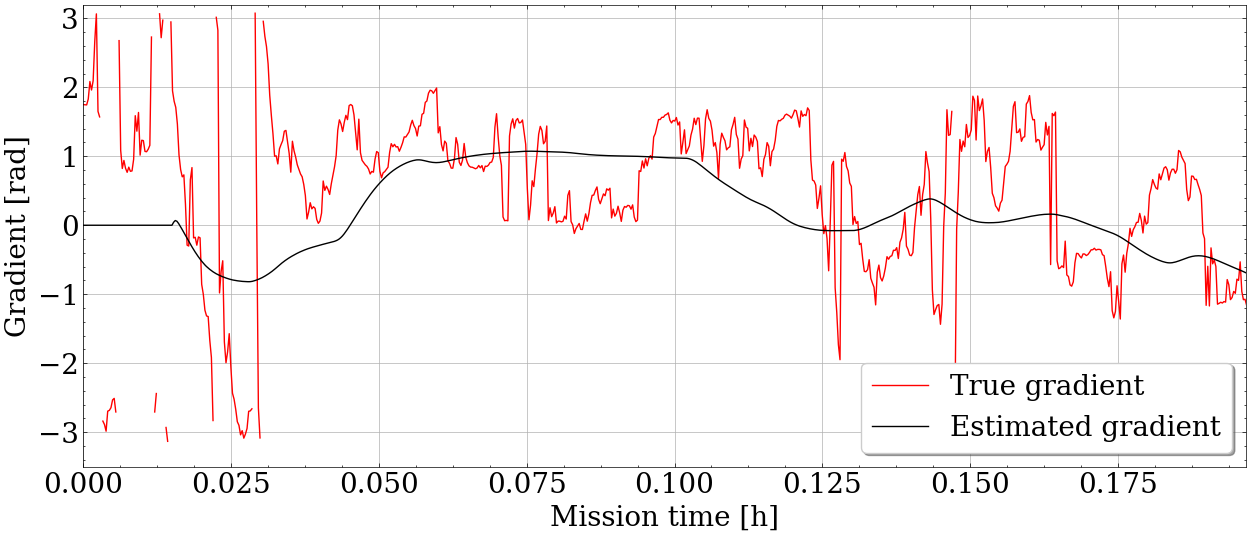}
      \vspace{-5mm}
      \caption{Survey in the 18th of July 2022.}
      \label{fig:experiment_July_grad}
    \end{subfigure}%
    \begin{subfigure}{.33\textwidth}
      \centering
      \includegraphics[width=\linewidth]{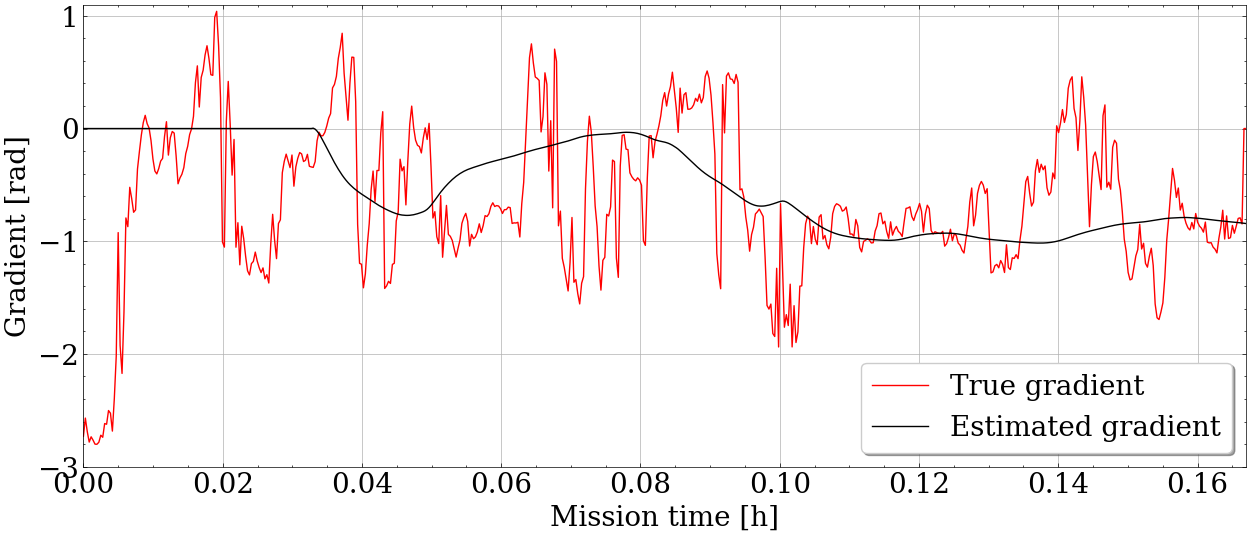}
      \vspace{-5mm}
      \caption{Survey in the 11th of August 2022.}
      \label{fig:experiment_Aug_grad}
    \end{subfigure}
    \begin{subfigure}{.33\textwidth}
      \centering
      \includegraphics[width=\linewidth]{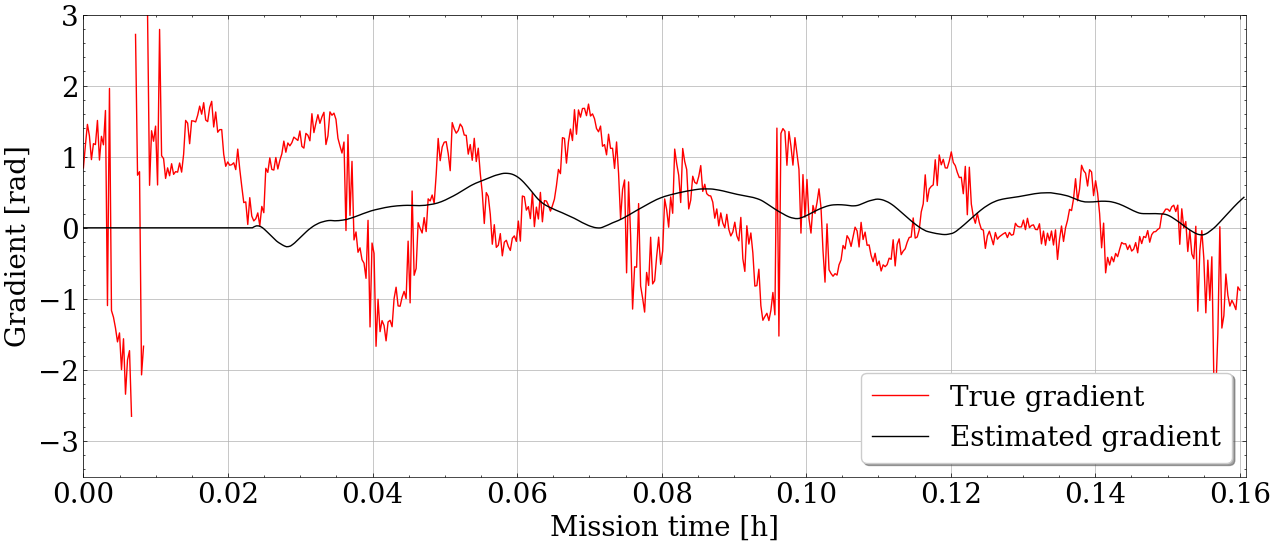}
      \vspace{-5mm}
      \caption{Simulation under experiment conditions.}
      \label{fig:scaled_grad}
    \end{subfigure}
     \vspace{-2mm}
     \caption{Gradient of chlorophyll a: AUV estimated gradient, and true gradient.}
    \label{fig:exp_grad}
    \vspace{-3mm}
\end{figure*}

We can also notice how the front tracking errors appear large for both surveys and the simulated scenario, as the AUV path resembles a sinusoid around the front.
In the previous section, we mentioned that the front tracking error in simulations is due to the AUV turning radius and the delay introduced by the update function of the gradient estimator.
Here, beyond those factors, we also have introduced a look-up table to simulate the sensor that consists of the {\Chl} concentration dataset, which has been scaled 100 fold. 
Considering that, if we compare the results of the previous section and the simulated scenario, the overshoot always tends to be approximately 10 meters.
We can further understand the implications of these overshoots by analyzing the {\Chl} concentration tracking error in Fig.~\ref{fig:exp_chl}.
Here, the overshoots and undershoots in {\Chl} concentration correspond to the overshoots and undershoots in the distance while tracking the front.
Bringing our attention to the gradient estimation, even with a jittery trajectory, the gradient error - the difference between the true gradient and estimated gradient - always seems below 90 degrees for all cases.
For a more detailed analysis, in Fig.~\ref{fig:exp_grad}, we can see how the estimated gradient resembles a low-pass filtered version of the true gradient with very high frequencies.

\begin{figure*}[htpb!]
    \centering
    \begin{subfigure}{.3\textwidth}
      \centering
      \includegraphics[width=\linewidth]{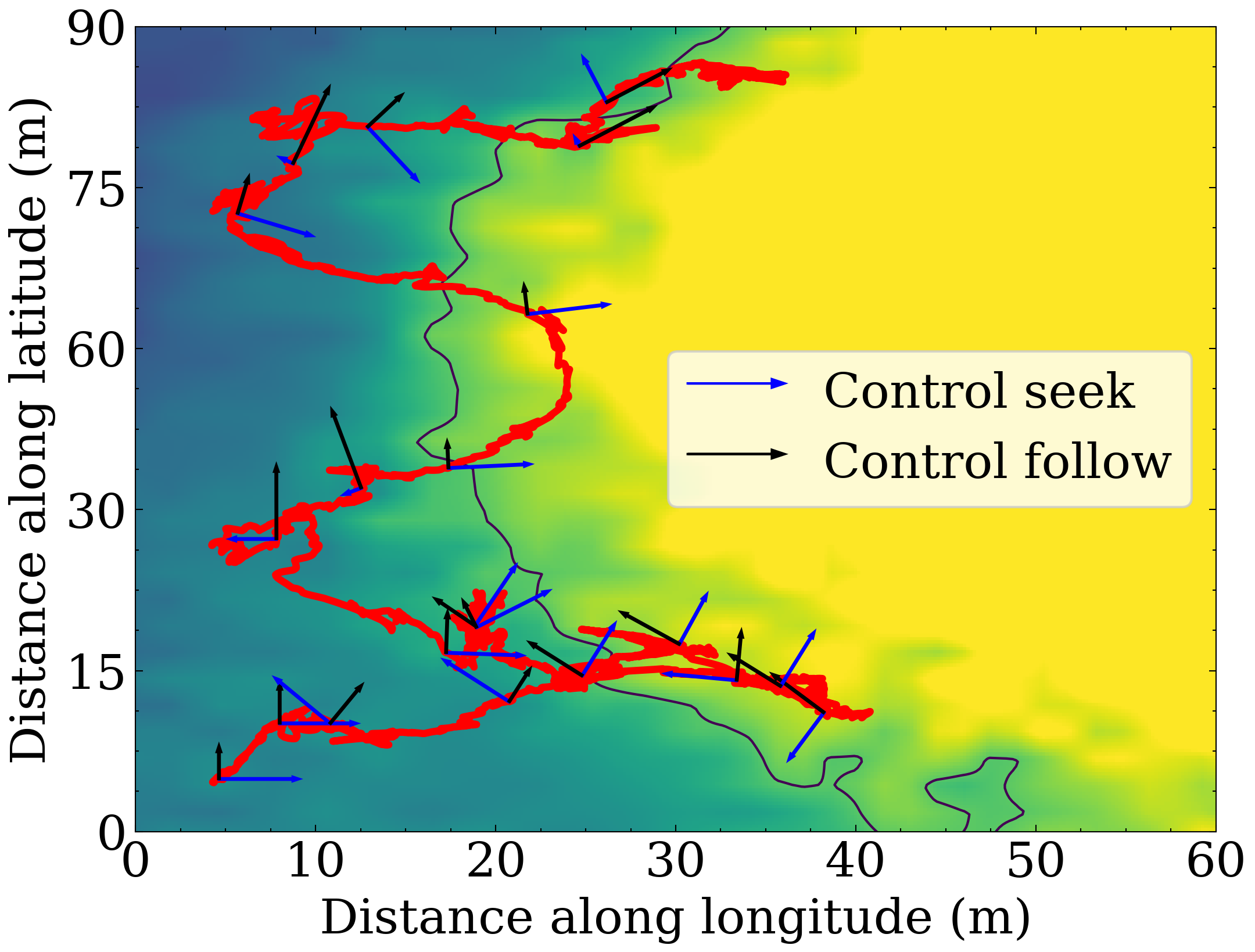}
      \vspace{-5mm}
      \caption{Survey in the 18th of July 2022.}
      \label{fig:experiment_July_zoom2}
    \end{subfigure}%
    \begin{subfigure}{.38\textwidth}
      \centering
      \includegraphics[width=\linewidth]{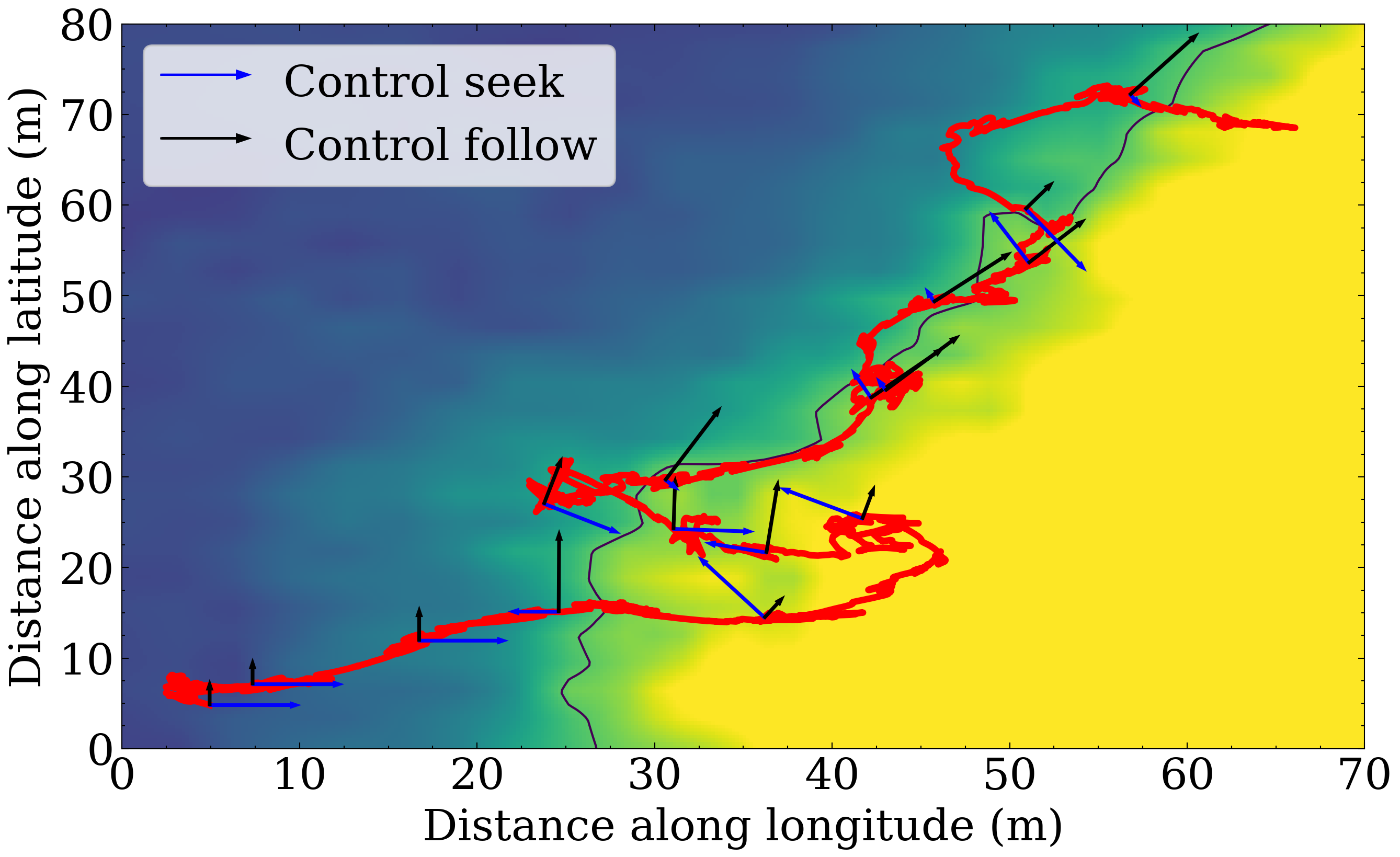}
      \vspace{-5mm}
      \caption{Survey in the 11th of August 2022.}
      \label{fig:experiment_Aug_zoom2}
    \end{subfigure}
    \begin{subfigure}{.3\textwidth}
      \centering
      \includegraphics[width=\linewidth]{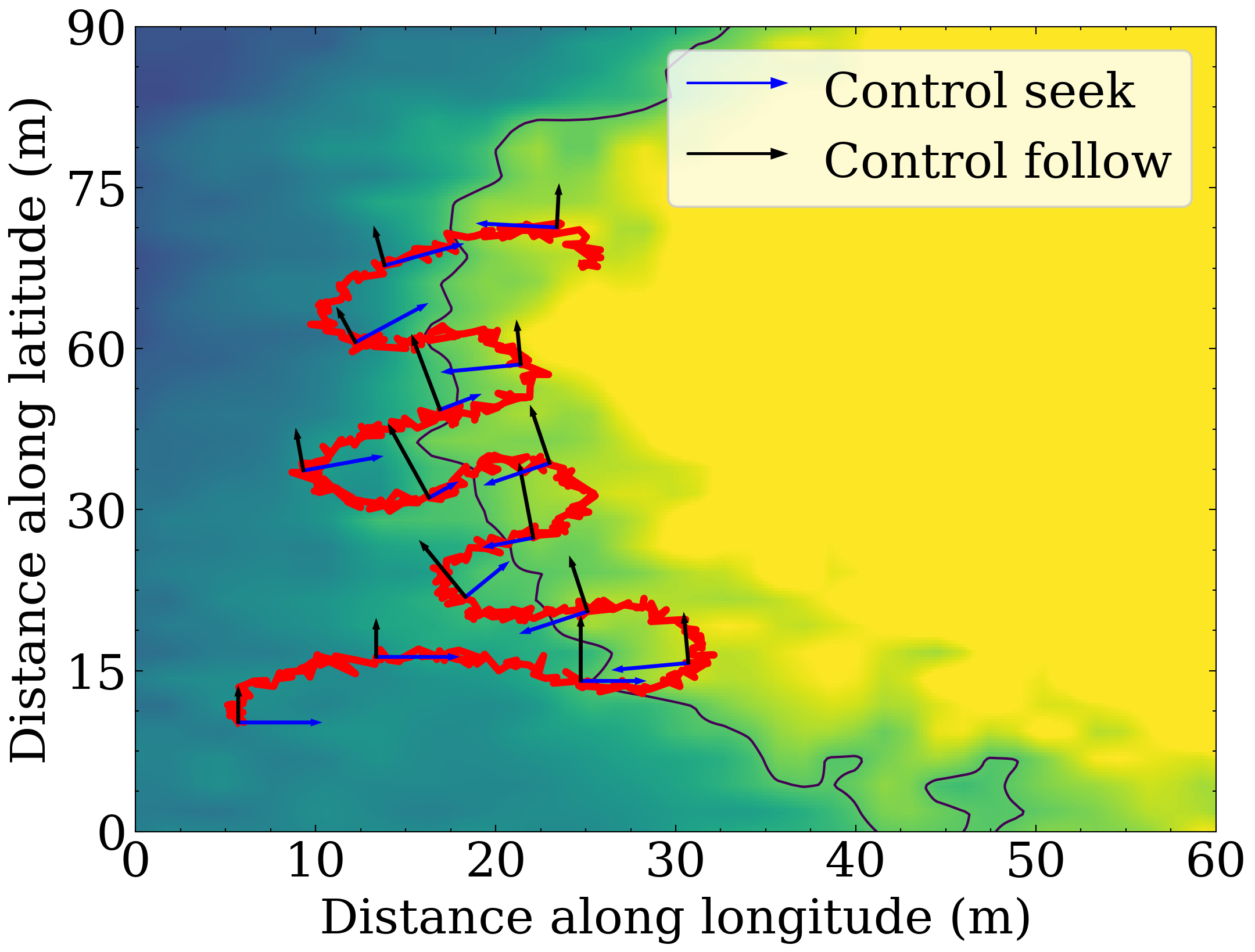}
      \vspace{-5mm}
      \caption{Simulation under experiment conditions.}
      \label{fig:scaled_zoom2}
    \end{subfigure}
     \vspace{-2mm}
     \caption{Trajectory of the AUV (red) tracking the front (black line), with arrows representing the true and estimated gradient.}
    \label{fig:exp_zoom2}
    \vspace{-3mm}
\end{figure*}

\begin{figure*}[htpb!]
    \centering
    \begin{subfigure}{.33\textwidth}
      \centering
      \includegraphics[width=\linewidth]{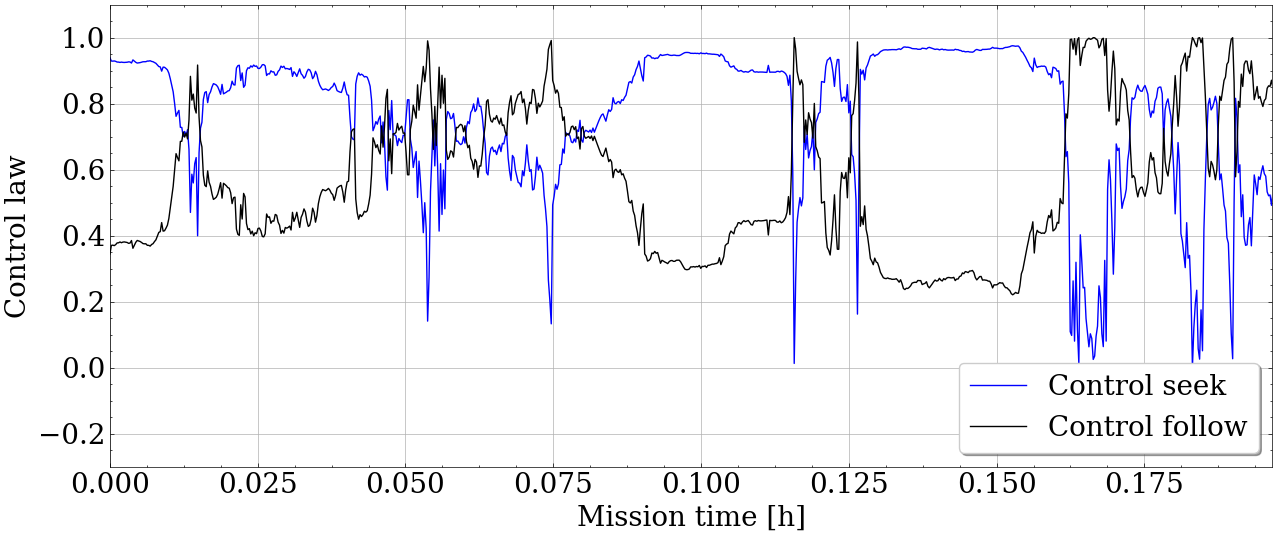}
      \vspace{-5mm}
      \caption{Survey in the 18th of July 2022.}
      \label{fig:experiment_July_control}
    \end{subfigure}%
    \begin{subfigure}{.33\textwidth}
      \centering
      \includegraphics[width=\linewidth]{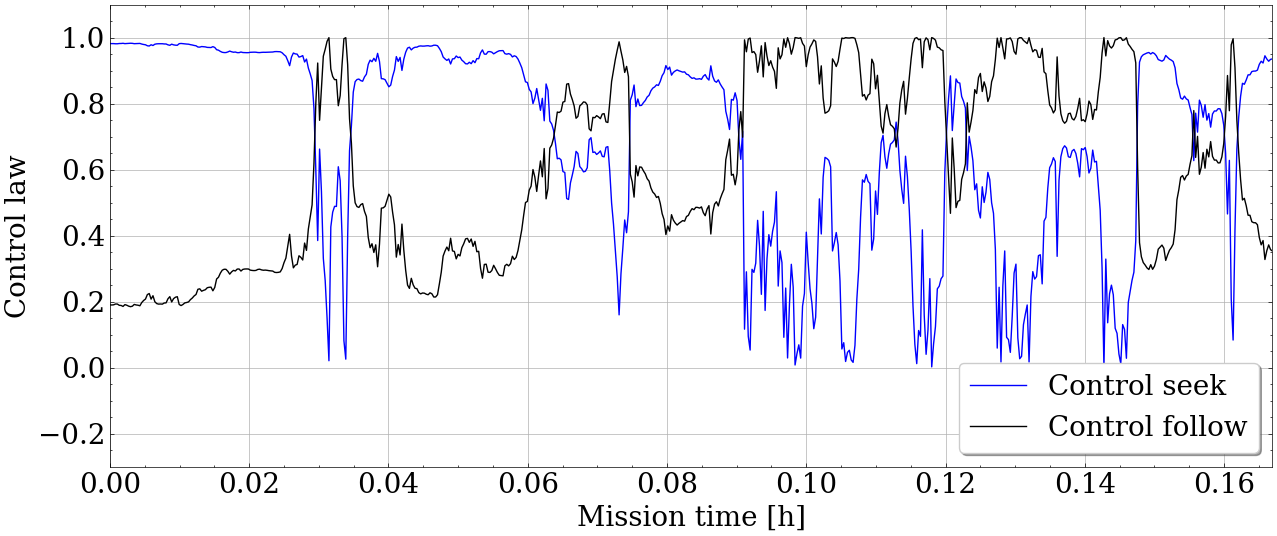}
      \vspace{-5mm}
      \caption{Survey in the 11th of August 2022.}
      \label{fig:experiment_Aug_control}
    \end{subfigure}
    \begin{subfigure}{.33\textwidth}
      \centering
      \includegraphics[width=\linewidth]{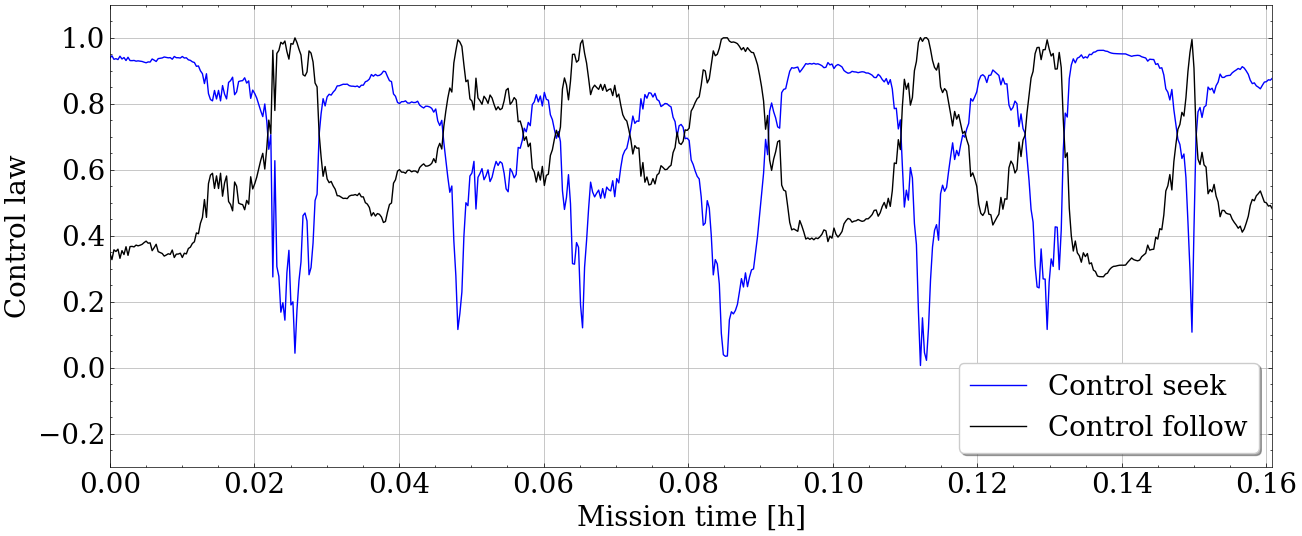}
      \vspace{-5mm}
      \caption{Simulation under experiment conditions.}
      \label{fig:scaled_control}
    \end{subfigure}
     \vspace{-2mm}
     \caption{Control law components: seek and follow.}
    \label{fig:exp_control}
    \vspace{-3mm}
\end{figure*}

In the second zoom-in in Fig.~\ref{fig:exp_zoom2}, we evaluate the performance of the control law and its components.
Here we note that the control law's seek and follow components match the jittery trajectory. 
The follow component tends to be parallel to the front, pointing forward, and the seek component tends towards the front.
Analysing Fig.~\ref{fig:exp_control}, we can see how the over and undershoots correspond to times where seek is the dominating component. 

\section{Conclusions and Future Work}\label{s: conclusions}

We considered the problem of how to use satellite data to improve adaptive sampling missions of an AUV equipped with a {\Chl} sensor and how to design a survey in the Baltic Sea to test our approach.
We developed the adaptive sampling algorithm and software packages to build a solution for the environmental sampling problem.
Our solution uses GPs to model {\Chl} fronts using satellite data and integrates such model into a front tracking algorithm. 
This integration is done using the estimate of the {\Chl} gradient field in the control law.
We confirmed the goodness of fit of the GP model by using scattered data points from a higher resolution satellite data and were able to reconstruct the {\Chl} field using the GP model.
We implemented the developed algorithm in the AUV's software and ran realistic simulations using the model of our AUV and {\Chl} sensor model.
These simulations resulted in accurate front tracking with low gradient estimation error.
We considered the two most important performance metrics for our objective to be gradient estimation errors and front tracking errors.
Concerning these metrics, the sensor noise analysis indicated that the gradient estimation using GP results in smaller errors than when using LSQ, mainly when the sensor noise is bigger or equal to $0.01\text{mg/m}^3$, which corresponds to most {\Chl} concentration sensors in the market.
We did not consider other performance metrics, such as time of computation, because both methods appeared fast enough to generate an estimate in real time.
We also did not consider the computation time for model fitting prior to the survey as it's not running in real-time and thus not affecting performance.
However, for some applications with fast-changing environments, it could be relevant to train the model during the survey. 
The GP estimator requires a GP model to be fit prior to the mission, which takes a few minutes, as opposed to the LSQ algorithm, which requires no prior fitting.
After the development, implementation, and sensitivity analysis, we designed a survey in the Baltic Sea, near Stockholm, next to the island of Djur{\"o}.
Similarly to the simulations, the experiments confirmed that the algorithm and software package work as desired in a controlled environment.

The work in this paper suggests several extensions, from theory to implementation.
Starting with more theoretical contributions, it should be noted that our approach was developed for front tracking. 
Though it cannot directly be used for other applications such as source tracking, zig-zagging around the front, or other movements, the control law can be adapted for such applications.
Another interesting extension to consider is the multi-agent scenario. 
This includes modifying the control law according to the neighbor agents and using the joint variance given by the GP model. 
Then the agents would be able to make more informed decisions regarding area coverage, knowing the areas with the highest variance.
A contribution that follows directly from multi-agent systems is data assimilation using the data collected by the multi-agent system.
This data could fit a GP model that informs algae population models through physics-informed learning.
On the implementation side, future work would include integrating the algae sensor into an AUV, tuning it in a controlled setting, and running experiments with the complete system.
One relevant contribution would be to work towards a higher degree of autonomy.
There are several steps to reach a level of robustness sufficient for real-world deployment of higher autonomy.
Some of them are reliable collision avoidance for islands, boats, and people, robust autonomous dock-in for charging, and a cloud-based data storage solution.

\section*{Acknowledgments}
We want to thank Josefine Severholt, Carl Ljung, Andreas Lezdins, and Koray Kulbay for their help with AUV hardware and support on the mission day, and Pedro Roque and Aldo Espinoza for their help with the AUV software, particularly with the control packages.
We would also like to thank CMEMS for the available satellite data.



\bibliographystyle{IEEEtran}
\bibliography{references}

\end{document}